\documentclass[twocolumn]{aastex701}

\hypersetup{linkcolor=red,citecolor=blue,filecolor=cyan,urlcolor=blue}
\usepackage{wrapfig}

\usepackage{xcolor}

\shorttitle{GRB250702B with JWST/NIRCam}
\shortauthors{Sears et al.}

\begin{document}

\author[0000-0001-8023-4912]{Huei Sears}
\email{huei.sears@rutgers.edu}
\affiliation{Department of Physics and Astronomy, Rutgers, the State University of New Jersey, 136 Frelinghuysen Road, Piscataway, NJ 08854-8019, USA}

\author[0000-0001-8426-5732]{Jean J. Somalwar}
\email{jsomalwar@berkeley.edu}
\affiliation{Department of Astronomy, University of California, Berkeley, CA 94720-3411, USA}
\affiliation{Berkeley Center for Multi-messenger Research on Astrophysical Transients and Outreach (Multi-RAPTOR), University of California, Berkeley, CA 94720-3411, USA}

\author[0000-0002-7706-5668]{Ryan Chornock}
\email{chornock@berkeley.edu}
\affiliation{Department of Astronomy, University of California, Berkeley, CA 94720-3411, USA}
\affiliation{Berkeley Center for Multi-messenger Research on Astrophysical Transients and Outreach (Multi-RAPTOR), University of California, Berkeley, CA 94720-3411, USA}

\author[0000-0003-1792-2338]{Tanmoy Laskar}
\email{tanmoy.laskar@utah.edu}
\affiliation{Department of Physics \& Astronomy, University of Utah, Salt Lake City, UT 84112, USA}

\author[0000-0001-7821-9369]{Andrew Levan}
\email{a.levan@astro.ru.nl}
\affiliation{Department of Astrophysics/IMAPP, Radboud University, 6525 AJ Nijmegen, The Netherlands}
\affiliation{Department of Physics, University of Warwick, Coventry, CV4 7AL, UK}

\author[0000-0003-4768-7586]{Raffaella Margutti}
\email{rmargutti@berkeley.edu}
\affiliation{Department of Astronomy, University of California, Berkeley, CA 94720-3411, USA}
\affiliation{Berkeley Center for Multi-messenger Research on Astrophysical Transients and Outreach (Multi-RAPTOR), University of California, Berkeley, CA 94720-3411, USA}
\affiliation{Department of Physics, University of California, 366 Physics North MC 7300, Berkeley, CA 94720, USA}

\author[0000-0002-9700-0036]{Brendan O'Connor}
\email{boconno2@andrew.cmu.edu}  
\altaffiliation{McWilliams Fellow}
\affiliation{McWilliams Center for Cosmology and Astrophysics, Department of Physics, Carnegie Mellon University, Pittsburgh, PA 15213, USA}

\author[0000-0002-8070-5400]{Nayana A. J.}
\email{nayana@berkeley.edu}
\affiliation{Department of Astronomy, University of California, Berkeley, CA 94720-3411, USA}
\affiliation{Berkeley Center for Multi-messenger Research on Astrophysical Transients and Outreach (Multi-RAPTOR), University of California, Berkeley, CA 94720-3411, USA}

\author[0000-0002-2184-6430]{Tomas Ahumada}
\email{tomas.ahumada@noirlab.edu} 
\affiliation{Division of Physics, Mathematics, and Astronomy, California Institute of Technology, Pasadena, CA 91125, USA}
\affiliation{Cerro Tololo Inter-American Observatory/NSF NOIRLab, Casilla 603, La Serena, Chile}
\author[0000-0002-8297-2473]{Kate~D.~Alexander}
\email{kdalexander@arizona.edu}
\affiliation{Department of Astronomy/Steward Observatory, 933 North Cherry Avenue, Rm. N204, Tucson, AZ 85721-0065, USA}

\author[0000-0002-8977-1498]{Igor Andreoni}
\email{igor.andreoni@unc.edu} 
\affiliation{Department of Physics and Astronomy, University of North Carolina at Chapel Hill, Chapel Hill, NC 27599-3255, USA}

\author[0000-0002-8935-9882]{Akash Anumarlapudi}
\email{akasha@unc.edu} 
\affiliation{Department of Physics and Astronomy, University of North Carolina at Chapel Hill, Chapel Hill, NC 27599-3255, USA}

\author[0000-0001-8544-584X]{Jonathan Carney}
\email{jcarney@unc.edu} 
\affiliation{Department of Physics and Astronomy, University of North Carolina at Chapel Hill, Chapel Hill, NC 27599-3255, USA}

\author[0009-0006-7990-0547]{James Freeburn}
\email{jfreebur@unc.edu} 
\affiliation{Department of Physics and Astronomy, University of North Carolina at Chapel Hill, Chapel Hill, NC 27599-3255, USA}

\author[0000-0002-1296-6887]{Llu\'is Galbany}
\email{lgalbany@ice.csic.es}
\affiliation{Institute of Space Sciences (ICE-CSIC), Campus UAB, Carrer de Can Magrans, s/n, E-08193 Barcelona, Spain}
\affiliation{Institut d'Estudis Espacials de Catalunya (IEEC), 08860 Castelldefels (Barcelona), Spain}

\author[0000-0002-5826-0548]{Benjamin P. Gompertz}
\email{b.gompertz@bham.ac.uk}
\affiliation{School of Physics and Astronomy, University of Birmingham, Edgbaston, Birmingham, B15 2TT, UK}
\affiliation{Institute for Gravitational Wave Astronomy, University of Birmingham, Edgbaston, Birmingham, B15 2TT, UK}

\author[0000-0002-4391-6137]{Or Graur}
\email{or.graur@port.ac.uk}
\affiliation{Institute of Cosmology \& Gravitation, University of Portsmouth, Dennis Sciama Building, Burnaby Road, Portsmouth PO1 3FX, UK}
\affiliation{Department of Astrophysics, American Museum of Natural History, New
York, NY 10024, USA}

\author[0000-0002-3841-380X]{Saarah Hall}
\email{saarah@u.northwestern.edu}
\affiliation{Department of Physics and Astronomy, Northwestern University, 2145 Sheridan Road, Evanston, IL 60208, USA}
\affiliation{Center for Interdisciplinary Exploration and Research in Astrophysics (CIERA), 1800 Sherman Ave., Evanston, IL 60201, USA}

\author[0000-0002-9364-5419]{Xander J. Hall}
\email{xhall@cmu.edu}
\affiliation{McWilliams Center for Cosmology and Astrophysics, Department of Physics, Carnegie Mellon University, Pittsburgh, PA 15213, USA}

\author[0000-0002-5698-8703]{Erica Hammerstein} 
\email{ekhammer@berkeley.edu}
\affiliation{Department of Astronomy, University of California, Berkeley, CA 94720-3411, USA}
\affiliation{Berkeley Center for Multi-messenger Research on Astrophysical Transients and Outreach (Multi-RAPTOR), University of California, Berkeley, CA 94720-3411, USA}

\author[0000-0001-8738-6011]{Saurabh W. Jha}
\email{saurabh@physics.rutgers.edu}
\affiliation{Department of Physics and Astronomy, Rutgers, the State University of New Jersey, 136 Frelinghuysen Road, Piscataway, NJ 08854-8019, USA}

\author[0000-0002-5619-4938]{Mansi M. Kasliwal}
\email{mansi@astro.caltech.edu} 
\affiliation{Division of Physics, Mathematics, and Astronomy, California Institute of Technology, Pasadena, CA 91125, USA}

\author[0000-0003-1386-7861]{Dheeraj Pasham}
\email{p.dheerajreddy@gmail.com} 
\affiliation{Eureka Scientific and George Washington University}

\author[0000-0003-0466-3779]{Itai Sfaradi}
\email{itai.sfaradi@berkeley.edu}
\affiliation{Department of Astronomy, University of California, Berkeley, CA 94720-3411, USA}
\affiliation{Berkeley Center for Multi-messenger Research on Astrophysical Transients and Outreach (Multi-RAPTOR), University of California, Berkeley, CA 94720-3411, USA}

\author[0000-0001-6747-8509]{Yuhan~Yao}
\email{yuhanyao@berkeley.edu}
\affiliation{Miller Institute for Basic Research in Science, 206B Stanley Hall, Berkeley, CA 94720, USA} 
\affiliation{Department of Astronomy, University of California, Berkeley, CA 94720-3411, USA}
\affiliation{Berkeley Center for Multi-messenger Research on Astrophysical Transients and Outreach (Multi-RAPTOR), University of California, Berkeley, CA 94720-3411, USA}

\title{Late-time JWST/NIRCam Observations of the Extremely Long-duration GRB 250702B/EP 250702a and Its Host Galaxy}

\correspondingauthor{Huei Sears}
\email{huei.sears@rutgers.edu}

\begin{abstract}

We present JWST/NIRCam observations of the extremely long-duration gamma-ray burst (GRB) 250702B taken at $\sim$95 days post-GRB (observer frame).  The observations of the host galaxy reveal a single galaxy with a prominent dust lane observed nearly edge-on. \texttt{Prospector} modeling of the host galaxy photometry finds a high stellar mass ($\log(M_*/M_{\odot}) = 11.0^{+0.2}_{-0.3}$) and large dust column ($A_V = 2.8 \pm 0.3$ mag), in agreement with previous results. If GRB 250702B is a collapsar-driven GRB, the host galaxy is the brightest (in rest-frame $r$ and rest-frame $H$) and most massive compared to GRB hosts at similar redshifts. The transient localization is near the dust lane, and while we find no evidence for transient emission in F277W, F356W, and F444W, forced photometry in F150W and F200W reveals possible $\sim 3\sigma$ detections of the transient at $m_{F150W} \sim 27.9$ AB mag and $m_{F200W} \sim 27.4$ AB mag. If these are secure detections, they are indicative of a late-time light curve flattening. This behavior is consistent with that of jetted tidal disruption events (TDEs); however, it is also consistent with a supernova plus GRB afterglow model.  Alternatively, if these are upper limits, they are consistent with, but do not further constrain, the extrapolated power-law decline of the afterglow.  The ambiguity of the possible detection of the transient in F150W and F200W highlights the need for late-time template observations with JWST/NIRCam.
\end{abstract}


\section{Introduction}
\label{sec:intro}

Gamma-ray bursts (GRBs) are known to be some of the most energetic phenomena in the Universe.  Observed phenomenologically as a burst of gamma-rays, these transients are often split by the duration of their gamma-ray emission (T90; the duration at which 90$\%$ of the emission is observed) into ``short" (T90 $<$ 2s, SGRB) and ``long" (T90 $>$ 2s, LGRB).  At least some SGRBs are known to arise from binary neutron star mergers (e.g., \citealt{MarguttiandChornock2021}), while LGRBs are known to arise from the core collapse of a massive star (e.g., \citealt{WoosleyandBloom2006}).  There are events which challenge this binary classification in progenitor path (e.g, GRBs 060505A \citep{Fynbo_2006, Ofek_2007}, 060614A \citep{Gehrels2006, DellaValle_060614, GalYam_060614, BingZhang_060614}, 111005A \citep{MichalowskI_111005A, Tanga_111005A, Wang_111005A}, 200826A \citep{Ahumada_200826A, Rossi_200826A, Zhang_200826A}, 211211A \citep{Rastinejad2022, Troja2022, Yang2022}, and 230307A \citep{Levan_230307A, Yang_230307A}), but overwhelmingly this classification otherwise observationally remains valid.

An additional subset of GRBs are the ``ultralong" (ULGRBs; \citealt{Virgili2013_grb091024A, Levan2014}), which have T90 $\gtrsim$ 1,000 s. There are a few-to-several ULGRBs known to the community \citep{goad2007_grb051117a, Virgili2013_grb091024A, Levan2014, Levan2015_ULGRBreview, Fu2024_grb211024b, Leung2026_grb220607a}. The exact classification requirements of a GRB into the ultralong class are still under debate, as long-lasting high-energy X-ray emission is sometimes considered part of the prompt emission \citep[e.g.,  GRB 121027A, ][]{Wu2013_GRB121027A} and longer T90 durations are sometimes required \citep[e.g., ][]{Burns2023}.  In addition, there are some ultralong GRBs which meet the T90 $\gtrsim$ 1,000 s duration requirement and are associated with the standard long GRB-supernova type (SN Ic-BL) however are intrinsically softer energy. These are classified as ``X-ray Flashes/XRFs" or ``low-luminosity GRBs" and are related to shock breakout (SBO) phenomena (e.g., GRBs 060218 (\citealt{Campana2006, Cobb2006, Modjaz2006, Pian2006, Sollerman2006, Wang2006}) and 100316D (\citealt{Bufano2011, Cano2011, Starling2011, Olivares2012, Bufano2012})). One of the first and, until recently, longest, ULGRB 111209A was also discovered to be associated with the bright supernova, SN 2011kl, which showed features between SNe Ic-BL and superluminous supernovae \citep[SLSNe,][]{Greiner2015, Kann2019}. The association of this SN was evidence that at least one ULGRB can be produced via similar core-collapse mechanisms as the standard LGRBs.

Another long-duration gamma-ray transient was Swift J1644+57 \citep{Bloom2011, Zauderer2011} which was classified as a tidal-disruption event (TDE) with a prompt on-axis jet (a ``jetted TDE").  In this scenario, a star is tidally-disrupted by a black hole and the accretion rate and environment is such that a relativistic jet can form.  There are only 4\footnote{A fifth jetted TDE candidate has been reported with EP 260302a (see JWST DD 12700, PI: Eyles-Ferris), however there are not yet any publications on this object.} known jetted TDEs: Swift J1644+57 \citep[$z$ = 0.354; ][]{Bloom2011, Zauderer2011, Levan2011}, Swift J2058+05 \citep[$z$ = 1.185; ][]{Cenko2012}, Swift J1112-82 \citep[$z$ = 0.890; ][]{Brown2015, Brown2017}, and AT 2022cmc \citep[$z$ = 1.193; ][]{Andreoni2022, Rhodes2023, Hammerstein2026}.  Three of these objects have locations consistent with the host galaxy nucleus, however no host galaxy has yet been detected for AT 2022cmc \citep{Andreoni2022, Hammerstein2026}. Based on these locations and numerous observational properties \citep[e.g., long-lived gamma-ray emission; bright, long-lived X-ray emission with $\sim$minute-timescale variability; radio synchrotron emission; ][]{Burrows2011, Bloom2011, Zauderer2011, Levan2011, Andreoni2022}, all are hypothesized to be tidal disruptions from supermassive black holes (SMBHs).  All  additionally show an abrupt X-ray decay at $\sim$100--400 rest-frame days after discovery following a power-law decay \citep{Eftekhari2024} and late-time optical/infrared (IR) plateaus \citep{ Brown2015, Pasham2015, Levan2016, Hammerstein2026}.

GRB 250702B was discovered on 02 July 2025 and triggered the Fermi Space Telescope several times with a total observed gamma-ray duration of T90 $\geq 25,000$ s \citep{Neights2026}.  The Einstein Probe also reported detections of both hard and soft X-ray emission up to 24 hours prior \citep[EP 250702a,][]{Li2026_EPpaperontheGRB}, leading to a ``prompt emission" duration of $\sim 110,000$ s, if accounting for this ``early" X-ray emission.  While no optical counterpart was found, due to the high extinction along the line-of-sight ($A_{V,\ MW} = 0.85$ mag, \citealt{Schlafly:2011}; $A_{V,\ host} = 5.77$ mag, \citealt{Levan2025_BDE, Gompertz2025}), a bright infrared source was detected at the transient location seemingly associated with an underlying extended source \citep{Levan2025_BDE}.  HST/WFC3 and JWST/NIRSpec observations of the field revealed this underlying extended source to be a galaxy at z = 1.036 $\pm$ 0.004 with the transient notably offset from the nucleus of the host galaxy \citep{Levan2025_BDE, Gompertz2025}.  Throughout this work, we use `GRB 250702B' to refer to this event.

The intrinsic nature of GRB 250702B is still an open question. The three main hypotheses for GRB 250702B are: 1) the standard LGRB collapsar process with extreme properties so as to allow the ``precursor" X-ray emission; 2) an off-nuclear, on-axis jetted TDE with $M_{BH} \leq 10^5 M_{\odot}$ \citep[across hypotheses 1 and 2,][]{Levan2025_BDE, Carney2025, OConnor2025, Gompertz2025, Beniamini2026,  Li2026_EPpaperontheGRB, Eyles-Ferris2026, ZhangJinPeng2026, Granot2026, Sato2026}; and 3) the merging of a black hole and a helium star, where the black hole accretion triggers core collapse \citep{Neights2026}.  These hypotheses offer testable predictions for late-time imaging: LGRBs are observed to have a power-law decay with a possible additional supernova component (which would appear simply as a bump atop this power-law decay), jetted TDEs have been observed to have a late-time UV/optical/near-infrared (NIR) plateau, and a helium-rich supernova is predicted for hypothesis three.

We present JWST/NIRCam imaging of the field of GRB 250702B in Section \ref{sec:observations}, along with previously unpublished NOT imaging.  In Section \ref{sec:photometry}, we measure host galaxy and transient photometry from the HST and JWST imaging. In Section \ref{sec:results}, we present a transient light curve and SED modeling of the host galaxy.  In Section \ref{sec:analysis}, we analyze the transient photometry in the context of jetted TDEs and ultralong GRBs, and we analyze the host photometry with that from off-nuclear TDE host galaxies and the broader long GRB host galaxy class.  Finally, in Section \ref{sec:conclusion}, we conclude the paper with insights from this object about the broader TDE and GRB classes. We adopt T$_{0,\text{ Fermi}}$ = 2025-07-02T13:56:05 as the start of the first Fermi trigger (`D') \citep{Neights2026}. We define times ($\Delta t$) in the observer frame with respect to this zeropoint, although the scientific conclusions from our late-time observations at $\sim 95$ d post-detection are not sensitive to whether we assume the Fermi or EP trigger for the zeropoint.  All magnitudes are reported on the AB system. We assume the WMAP9 cosmology with $H_0 = 69.33\ \text{km s}^{-1} \text{ Mpc}^{-1}$ and $\Omega_{c} = 0.2408$ \citep{Hinshaw2013_WMAP9}.

\section{Observations}
\label{sec:observations}

\subsection{JWST Imaging}
\label{subsec:JWSTimaging}
We present JWST/NIRCam imaging (ID: DD 9447; PI: H. Sears) from 05 October 2025 (F150W, F200W, F277W, F356W, and F444W; $\Delta$t = $+$94.60, 94.70, 94.57, 94.63, and 94.70  days, respectively).  The F150W imaging has an exposure duration of 10221 s, while F200W, F277W, F356W, and F444W each have exposure durations of 5110 s. For analysis in this work, we use the Level 3 products from the MAST archive. These observations are shown in a multi-color composite image in Figure \ref{fig:stscirelease}. The field is crowded with many foreground stars. The host galaxy is red and viewed edge-on with a distorted disk and a prominent dust lane across the nucleus in the bluest filters. GRB 250702B is located within the dust lane and is offset $\sim$5.5 kpc from the nucleus toward the south-east. 

\subsection{Hubble Space Telescope Imaging}
We re-analyze HST/WFC3 F160W imaging (ID: GO/DD 17988; \citealt{Levan_HSTproposal}) of GRB 250702B acquired at $\Delta$t = $+12.56$\,d, previously analyzed by \cite{Levan2025_BDE} and \cite{Carney2025}.  We set \texttt{final\_scale} = 0.065'' in the \texttt{AstroDrizzle} reduction step. The HST and JWST data used in this paper can be found in MAST: \dataset[10.17909/m93t-nc65]{http://dx.doi.org/10.17909/m93t-nc65}. 

\subsection{Nordic Optical Telescope (NOT) Imaging}

We obtained NOT/NOTCam $J$ imaging of the field of GRB 250702B on 08 September 2025 for a total exposure time of 1560 seconds. We report a non-detection of both the host and the transient. Using a Vista Hemisphere Survey image of the field for normalization and custom \texttt{astropy} routines, we report a $3\sigma$ detection limit of $m_{J,\ \text{Host}} > 18.4$ AB mag. Due to the non-detection of both the host and the transient, we do not include this limit in further analysis.

\begin{figure*}
\centering
\includegraphics[width = 1 \textwidth]{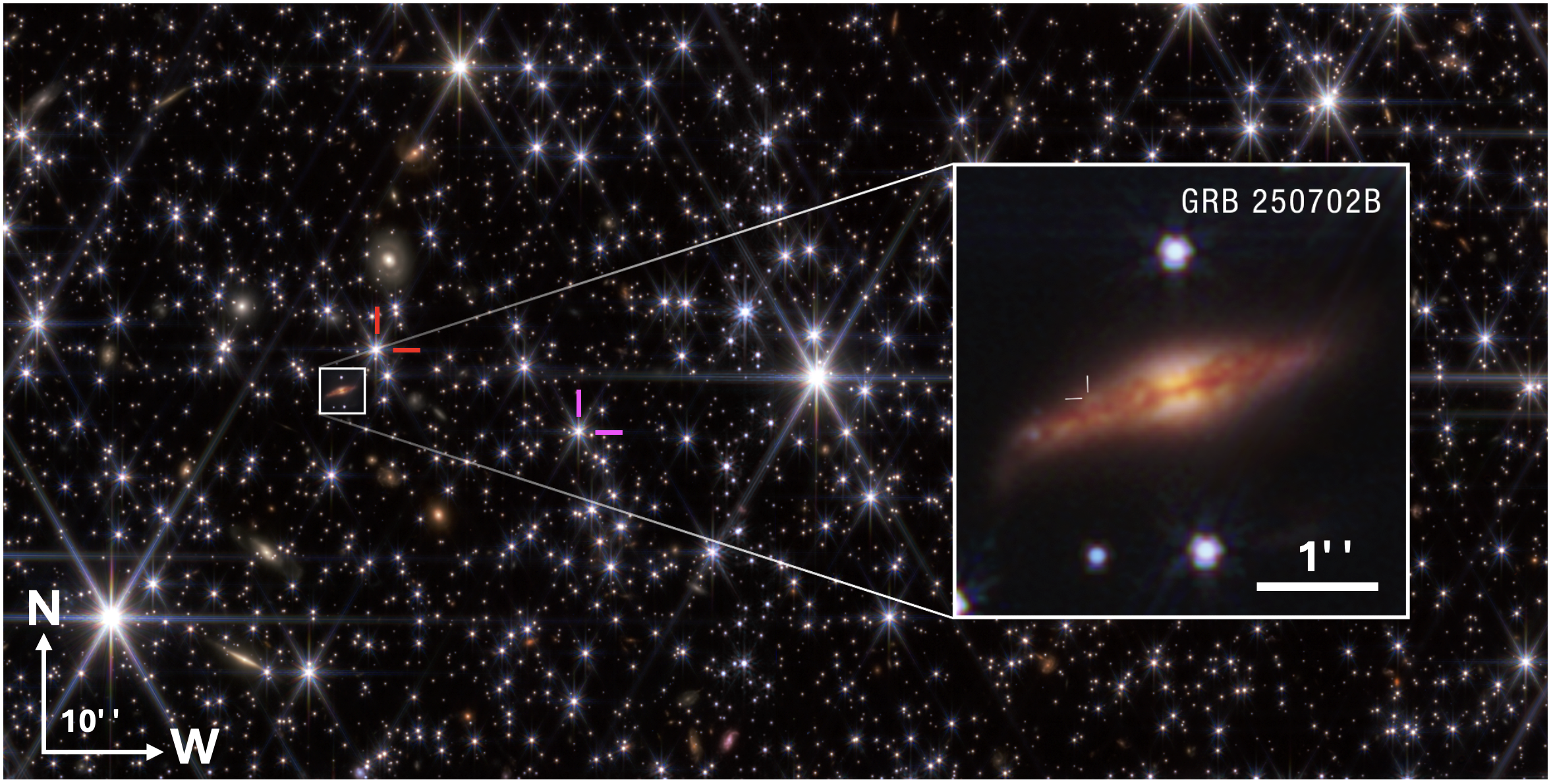}
\caption{A multi-color JWST/NIRCam composite image of the field and host galaxy of GRB 250702B. The zoomed inset shows the host galaxy of GRB 250702B, which shows a large, dusty, spiral galaxy. The transient is marked with white cross hairs in the zoomed inset, and it is offset by $\sim 0.67''$ ($\sim 5.5$ kpc at $z = 1.036$) from the nucleus of the host galaxy. The contaminating star and the subtraction star are marked with red and magenta cross hairs, respectively. The diffraction spikes near the host galaxy have been removed to better see host galaxy features, but the original field in F150W is shown in Figure \ref{fig:diffspike_sub}. The imaging used in this figure is described in Section \ref{subsec:JWSTimaging}. This figure has been modified from its originally presented version. Credit: NASA, ESA, CSA, H. Sears (Rutgers). Image Processing: A. Pagan (STScI).\label{fig:stscirelease}}
\end{figure*}

\section{Photometry}
\label{sec:photometry}

\subsection{Diffraction Spike Subtraction}
\label{subsec:diffspike}
\begin{figure*}
\includegraphics[width = \textwidth]{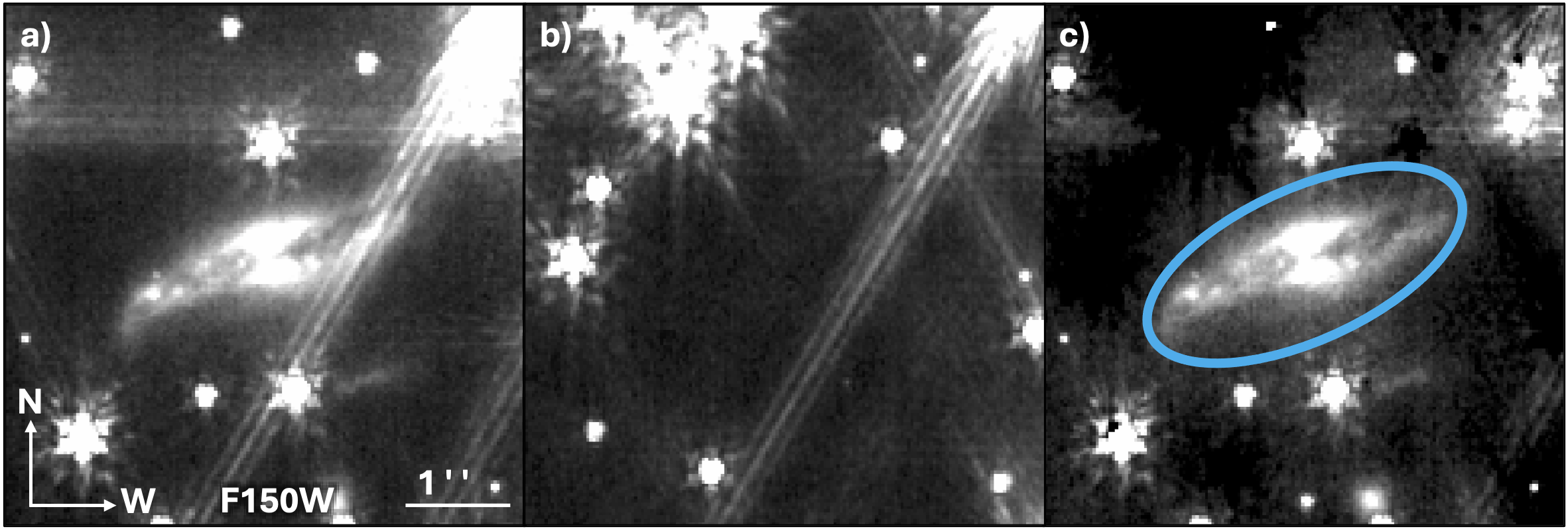}
\caption{Steps showing the subtraction of the diffraction spike contaminating the host galaxy in F150W. The pixel scale is the same across panels. Panel a) shows the original field around the host galaxy of GRB 250702B. Panel b) shows another diffraction spike in the image which is used for subtraction. Panel c) shows the subtraction of panel b) from panel a).  The blue ellipse shows the extraction region used for host galaxy aperture photometry. This procedure is described in Section \ref{subsec:diffspike}. \label{fig:diffspike_sub}}
\end{figure*}

GRB 250702B is located at a Galactic latitude of $b = -5^\circ$ \citep{Levan2025_BDE}, near the Galactic plane. As such, the field around the host galaxy of GRB 250702B is crowded by stars from the Milky Way, and the JWST host galaxy photometry is contaminated by a diffraction spike from a nearby star (see Figure \ref{fig:diffspike_sub}).  To remove this contamination, we subtract the diffraction spike of a similarly bright star (marked with magenta cross hairs in Figure \ref{fig:stscirelease}) with no sources underlying where the host galaxy would be in the respective frame.  To account for any variability between differently-angled spikes, we ensure that the spike we are subtracting is at the same orientation of the offending spike on the host galaxy. We first create cropped images of both the host galaxy of GRB 250702B and the diffraction spike of the template star. We then sub-pixel align both images, using custom routines in Python and visual inspection. Finally, we subtract the template star image from that of GRB 250702B. We find that we have to scale the second star by a factor of 1.2 to achieve a subtraction with no visual residual of the diffraction spike.  We repeat this process for each filter and find the 1.2 scaling appropriate in all filters.  In the same process, we also create error images from the `\texttt{\_err}' frames of the Level 3 `\texttt{\_i2d}' data products. The `\texttt{\_err}' frames contain resampled uncertainty estimates and are given as standard deviations.\footnote{\url{https://jwst-pipeline.readthedocs.io/en/latest/jwst/data_products/science_products.html\#i2d}} We repeat the same alignment and scale process as for the subtracted science frames; however, we instead add the error in quadrature. This process is shown in Figure \ref{fig:diffspike_sub} for F150W.  We perform aperture photometry of the host galaxy of GRB 250702B pre- and post-diffraction-spike subtraction, and we measure a 0.01 mag change in all filters except F150W, where we measure a 0.05 mag change.  All further analysis in this work uses the diffraction-spike subtracted images.

\subsection{JWST Transient Photometry}
\label{subsec:jwsttransient}

To determine the location of the transient, using custom Python routines, we perform relative astrometry between the NIRCam imaging to earlier HAWK-I imaging presented in \cite{Levan2025_BDE}, where the transient is brighter, utilizing $\sim 20$ stars in the HAWKI image that are not significantly saturated in the NIRCAM observations.  We report the location of the transient as RA: 18$^{\text{h}}$ 58$^{\text{m}}$  45$^{\text{s}}$.567, Dec: $-07^\circ$ 52\arcmin\ 26\arcsec .265 in the ICRS coordinates automatically assigned to the Level 3 products. The relative astrometric accuracy of this position is 0.015\arcsec\ based on the scatter in the geometric transformation between HAWKI and NIRCAM frames, while the absolute position is subject to additional uncertainty of the alignment of the NIRCAM images to the Gaia frame, although we note with the low Galactic latitude, the number of sources for this match is large.

\begin{figure*}
\centering
\includegraphics[width = 1 \textwidth]{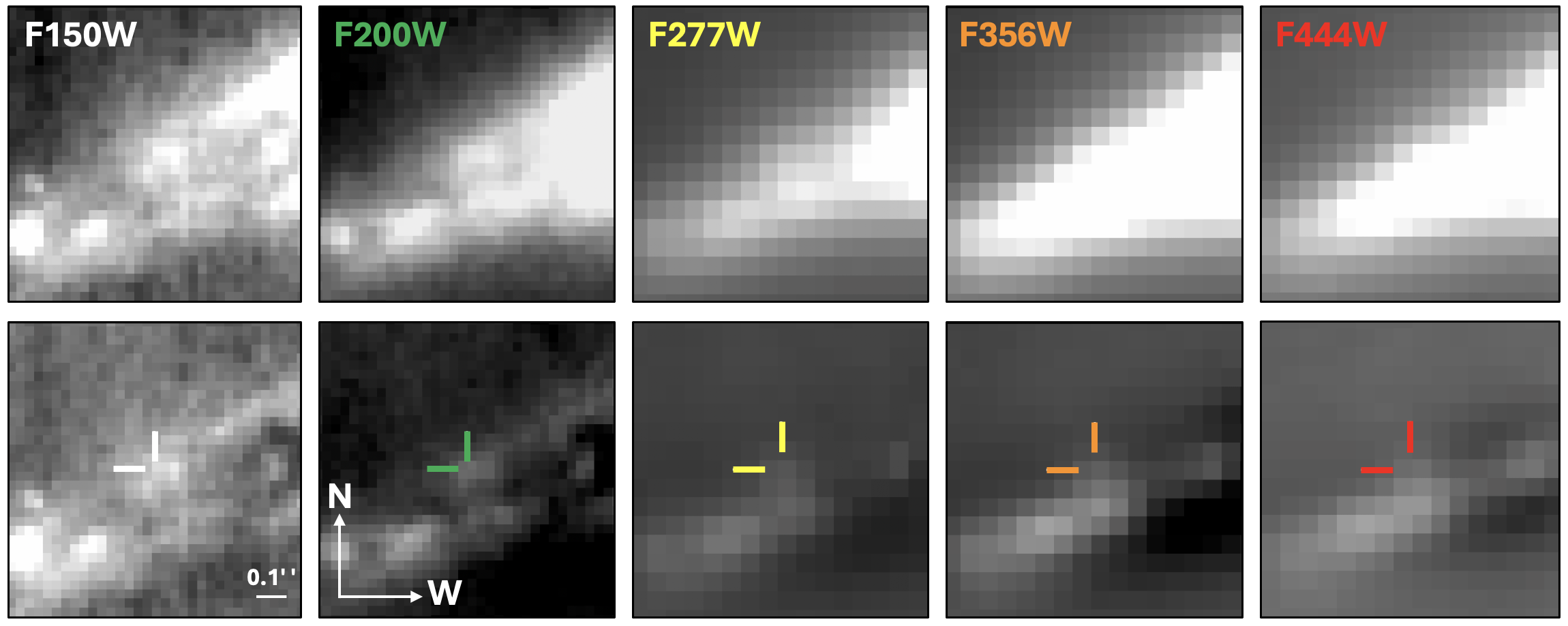}
\caption{1\arcsec\ crops of the field around the transient location. The top row shows the Level 3 data products, and the bottom row shows the \texttt{GALFIT} residual after galaxy and sky subtraction.  The cross hairs show the location of the transient.  The scale and colorbar have been adjusted to see fine details. \label{fig:zoomins}}
\end{figure*}

The host galaxy of GRB 250702B has a complex morphology and displays a prominent dust lane, which covers the location of the transient.  To isolate emission from the transient, we subtract most of the light of the galaxy using \texttt{GALFIT} \citep{Peng2010} with the assumption of a Sérsic component for the nucleus and an \texttt{edgedisk} profile for the disk.  For F150W, F200W, and F277W, where the dust lane appears to divide the nucleus into two parts, we assume two Sérsic profiles for the nucleus.  We fix the sky to be a constant at a value we measure in a source-free portion of the image. For the input point-spread function (PSF) for \texttt{GALFIT}, we use the \texttt{in\_flight\_opd} PSFs for NIRCam from the `Library of Simulated PSFs.'\footnote{\url{https://stsci.app.box.com/v/jwst-simulated-psf-library}}  We additionally use the summed `\texttt{\_err}' image for the \texttt{GALFIT} sigma image. We show a 1\arcsec\ crop of the location of the transient in the science and residual frames after the \texttt{GALFIT} subtraction in Figure \ref{fig:zoomins}. We find a complex residual, as expected from the science frame, and while there appears to be flux at the location of the transient, it is not obvious how much of this is attributable to the transient or to clumpy structure (e.g., star-formation, dust, galactic structure, etc.) along the line-of-sight. We discuss implications of this ambiguity in later sections.

To determine the 3$\sigma$ detection limit against the background of the host galaxy, we perform repeated forced photometry on the \texttt{GALFIT} residual with a circular aperture of radius equal to the FWHM of the PSF at locations separated by the FWHM.  For these locations, we exclude the nucleus, the location of the transient, and the extension on the south-east side of the galaxy.  For F150W, we also exclude an elliptical region around the dark dust lane on the west side of the galaxy (shown in Figure \ref{fig:diffspike_sub}). We make 461, 226, 280, 70, and 48 measurements in F150W, F200W, F277W, F356W, and F444W, respectively.  The differences in numbers of measurements are due to the differences in pixel scale between the short-wavelength and long-wavelength filters.  We report our $3\sigma$ limit as 3 times the standard deviation of the forced photometry across all measurements.  These limits are reported in Table \ref{tbl:transientphot}.

To measure flux at the location of the transient, we perform forced aperture photometry and forced PSF photometry on the \texttt{GALFIT} residual.  For the forced aperture photometry, we assume a circular aperture with a radius of the PSF FWHM.  We use the FWHM and encircled energy corrections as reported in the `NIRCam Point Spread Functions' section of the JWST User Documentation.\footnote{\url{https://jwst-docs.stsci.edu/jwst-near-infrared-camera/nircam-performance/nircam-point-spread-functions\#gsc.tab=0}}   For the forced aperture photometry uncertainty, we adopt the standard deviation of the flux measured from the methods used above to estimate the detection limit.  We choose not to incorporate the uncertainty from the summed `\texttt{\_err}' image in the forced aperture photometry uncertainty calculation, as the uncertainty introduced by the variable host galaxy background dominates over the statistical uncertainty introduced by the instrument and reduction steps. For the forced PSF photometry, we rerun \texttt{GALFIT} with an additional PSF object at the location of the transient.  We hold all galaxy and sky parameters fixed to the previous best-fit values and only allow the magnitude of the PSF for the transient to be free.  These measurements are reported in Table \ref{tbl:transientphot}.

\begin{deluxetable}{lrrrrr}
    \tablecaption{JWST + HST Transient Photometry of GRB 250702B. The forced aperture photometry uncertainty assumes a flux density uncertainty at the 1$\sigma$ confidence level (c.l.).  If the flux density uncertainty is greater than the measured forced aperture flux, there is no magnitude uncertainty listed. The uncertainty for the forced \texttt{GALFIT} photometry is that reported by \texttt{GALFIT}.  Details about these measurements are in Sections \ref{subsec:jwsttransient} and \ref{subsec:hsttransient}. \label{tbl:transientphot}}
    \tablehead{\textbf{Filter} & \textbf{$\Delta t$} & \textbf{1$\sigma$ c.l.} & \textbf{3$\sigma$ Limit} & \textbf{Forced Aperture Photometry} & \textbf{\texttt{GALFIT} PSF}\\ 
    & \textbf{[d]}& \textbf{[mJy]} & \textbf{[AB mag]}& \textbf{[AB mag]} & \textbf{[AB mag]}}
    \startdata
        JWST/F150W & 94.57 & $8.23 \times 10^{-6}$ & $27.91$ & $27.92^{+0.44}_{-0.31}$ & $28.06 \pm 0.14$ \\
        JWST/F200W & 94.67 & $2.03 \times 10^{-5}$ & $26.94$ & $27.38^{+0.75}_{-0.44}$ & $27.30 \pm 0.12$ \\
        JWST/F277W & 94.54 & $5.08 \times 10^{-5}$ & $25.94$ & $28.20$ & $26.82 \pm 0.29$ \\
        JWST/F356W & 94.60 & $1.51 \times 10^{-4}$ & $24.75$ & $27.48$ & $26.18 \pm 0.26$ \\
        JWST/F444W & 94.67 &  $1.65 \times 10^{-4}$ & $24.68$ & $27.38$ & $26.08 \pm 0.22$ \\
        HST/F160W & 12.53 & - & - & $25.98 \pm 0.07$ & - \\
    \enddata
\end{deluxetable}

\subsection{JWST Host Galaxy Photometry}
\label{subsec:jwsthost}
Given the lack of a bright detection of the transient in any filter, we perform aperture photometry on the host galaxy.  We use an elliptical aperture in all filters with a semi-major and semi-minor axis of 1.631\arcsec\ and 0.721\arcsec, respectively.  This aperture was selected based on visual inspection of the JWST imaging and is similar to that used in \cite{Levan2025_BDE} (semi-major/minor axes 1.75\arcsec/0.65\arcsec).  We subtract the sky and perform aperture photometry with \texttt{Photutils} \citep{larry_bradley_2025}.  We report an uncertainty of 0.05 mag on all measurements to account for both the uncertainty propagated from the error frame and encircled energy corrections on the order of $\sim 5\%$ (following the JWST User Documentation).  We report these measurements in Table \ref{tbl:JWSThostphot}.

\begin{deluxetable}{lr}
    \tablecaption{JWST photometry of the host galaxy of GRB 250702B. Methods used to make these measurements are described in Section \ref{subsec:jwsthost}. These values are not corrected for Galactic extinction. \label{tbl:JWSThostphot}}
    \tablehead{\textbf{Filter} & \textbf{Host Magnitude}\\
    & \textbf{[AB mag]}}
    \startdata
        F150W & $21.38 \pm 0.05$ \\
        F200W & $20.64 \pm 0.05$ \\
        F277W & $20.04 \pm 0.05$ \\
        F356W & $19.63 \pm 0.05$ \\
        F444W & $19.64 \pm 0.05$ \\
    \enddata
\end{deluxetable}

\subsection{HST Transient Photometry}
\label{subsec:hsttransient}
\begin{figure*}
\centering
\includegraphics[width = 1 \textwidth]{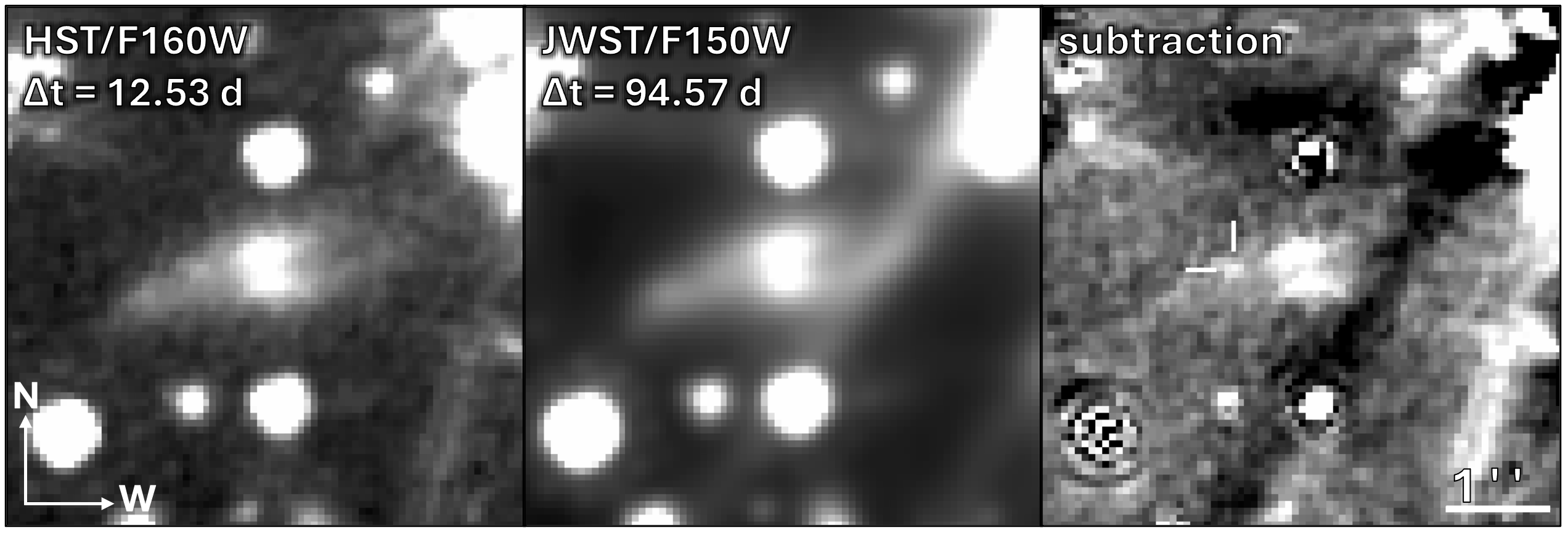}
\caption{Steps showing the subtraction of the JWST/F150W image from the HST/F160W image.  The left panel shows the original field around the host galaxy of GRB 250702B in HST/F160W.  The pixel scale is the same across all panels. The middle panel shows the same field in the JWST/F150W image. The image has been rescaled to the pixel scale of the HST image and has also been convolved with the HST/F160W point-spread function. The right panel shows the subtraction of the JWST/F150W image from the HST/F160W image. There is a clear detection of flux at the location of the transient (marked with white crosshairs). Methods used to make this figure are described in Section \ref{subsec:hsttransient}. \label{fig:template_sub}}
\end{figure*}

Given the possible lack of evidence for transient flux in F150W, we use the JWST/NIRCam F150W image as a template for subtraction for the earlier ($\Delta t$=12.53~d) HST/WFC3 F160W image.  We first re-grid the JWST Level 3 products to the plate scale of the drizzled HST image (0.065\arcsec). To do this step, we use \texttt{swarp} \citep{Bertin2002} with \texttt{RESAMPLING TYPE} set to `LANCZOS3.' We then construct an observed PSF in the HST image, which we use to convolve with the NIRCam F150W image.  To measure the PSF, we use a custom routine including steps involving the use of \texttt{Source Extractor} \citep{Bertin1996} and the \texttt{astropy} package \texttt{EPSFBuilder} \citep{larry_bradley_2025}, as similarly used in \cite{Sears2025}.  We convolve the re-sampled JWST F150W image using the \texttt{convolve\_fft} function within the \texttt{astropy.convolution} package \citep{larry_bradley_2025}. Finally we subtract the re-sampled, convolved F150W image from the F160W image using \texttt{hotpants} \citep{Becker2015}.  We force the convolution (\texttt{-c}) on the template, normalize (\texttt{-n}) to the image, assume a kernel order (\texttt{-ko}) of 1, and assume a background order (\texttt{-bgo}) of 0.5.  We visually detect a source at the location of the transient (see Figure \ref{fig:template_sub}), and we complete forced photometry in a 0.2\arcsec-radius aperture.  We report uncertainty on this measurement as propagation of uncertainty from the measurement of the sky.  We report an apparent magnitude of $m_{F160W} = 25.98 \pm 0.07$ AB mag. We separately repeat the subtraction step and photometry measurement without prior convolving the re-sampled JWST image (since \texttt{hotpants} runs its own convolution), and we find our result is not sensitive to this choice.  We also repeat all steps with a larger crop (so as to aid \texttt{hotpants} in the normalization step), and we similarly find our result is not sensitive to this choice. 

\section{Results}
\label{sec:results}

\subsection{Transient Light Curve}
\label{subsec:lightcurve}

\begin{figure}
\centering
\includegraphics[width = 1\columnwidth]{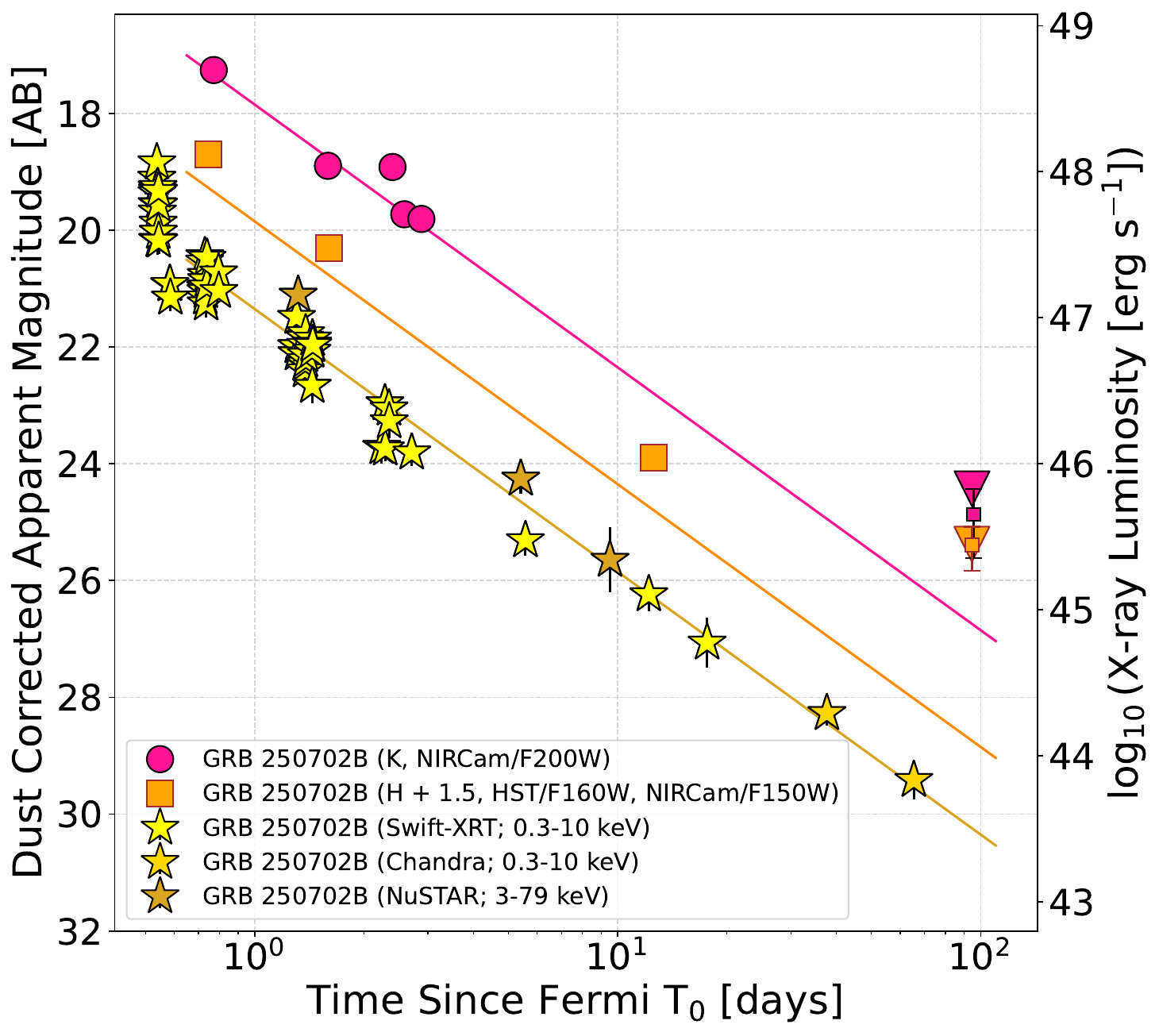}
\caption{The IR and X-ray light curves of GRB 250702B. Circles (magenta) and squares (orange) show IR detections, while the similarly-colored downward pointing triangles (smaller squares) show JWST upper-limits (forced aperture photometry).  Stars (shades of yellow) show X-ray detections from the listed observatories. The solid line extrapolations all have a slope of $t^{-1.8}$ (the best-fit X-ray slope from \citealt{OConnor2025}) and are arbitrarily offset to match the three GRB 250702B data sets. Uncertainty bars are sometimes smaller than the symbols.  See Table \ref{tbl:transientphot} for all IR measurements of GRB 250702B used in this plot. X-ray observations of GRB 250702B are compiled from the Swift-XRT archive and Table 1 in \cite{OConnor2025}. \label{fig:lightcurve}}
\end{figure}

GRB 250702B was well-detected in $H$ and $K$ in observations at $\Delta$t  $\sim$ 0.7--13 d \citep{Levan2025_BDE, Carney2025}.  We compile the published photometry with the newly measured JWST and HST photometry presented in this work (Figure \ref{fig:lightcurve}).  We correct all photometry for Galactic extinction, $A_{V,\ MW} = 0.85$ mag \citep{Schlafly:2011}, and host extinction, $A_{V,\ host} = 5.77$ mag \citep{Levan2025_BDE, Gompertz2025} using \texttt{dust\_extinction} \citep{Gordon2024} with the \texttt{G23} model \citep{Gordon2009, Fitzpatrick2019, Gordon2021, Decleir2022, Gordon2023}. In a study of X-ray photometry spanning $\Delta$t  $\sim$ 0.5--70 d, \cite{OConnor2025} find a power-law decay of the flux of $t^{-1.8}$ (see also Figure \ref{fig:lightcurve}). We find agreement with the $t^{-1.8}$ decay with the published $H$, $K$, and our re-measured HST observations, but the JWST limits are non-constraining.  Further investigation of these data, including comparison to other objects, is presented in Section \ref{sec:analysis}.

\subsection{Host SED Modeling}
\label{subsec:Proscpector}

We model the host galaxy photometry to extract the physical parameters of the stellar population, as well as their posterior distributions. We use the \texttt{Prospector} \citep{Leja2017, Johnson2019, Johnson2021} package to enable a full Bayesian statistical inference. \texttt{Prospector} computes stellar population models using the \texttt{Python-FSPS} \citep[][]{JohnsonDfm/python-fsps:V0.4.7} wrapper of the \texttt{FSPS} stellar population synthesis code \citep{Conroy2010FSPS:Synthesis}.

\begin{figure}
\centering
\includegraphics[width = \columnwidth]{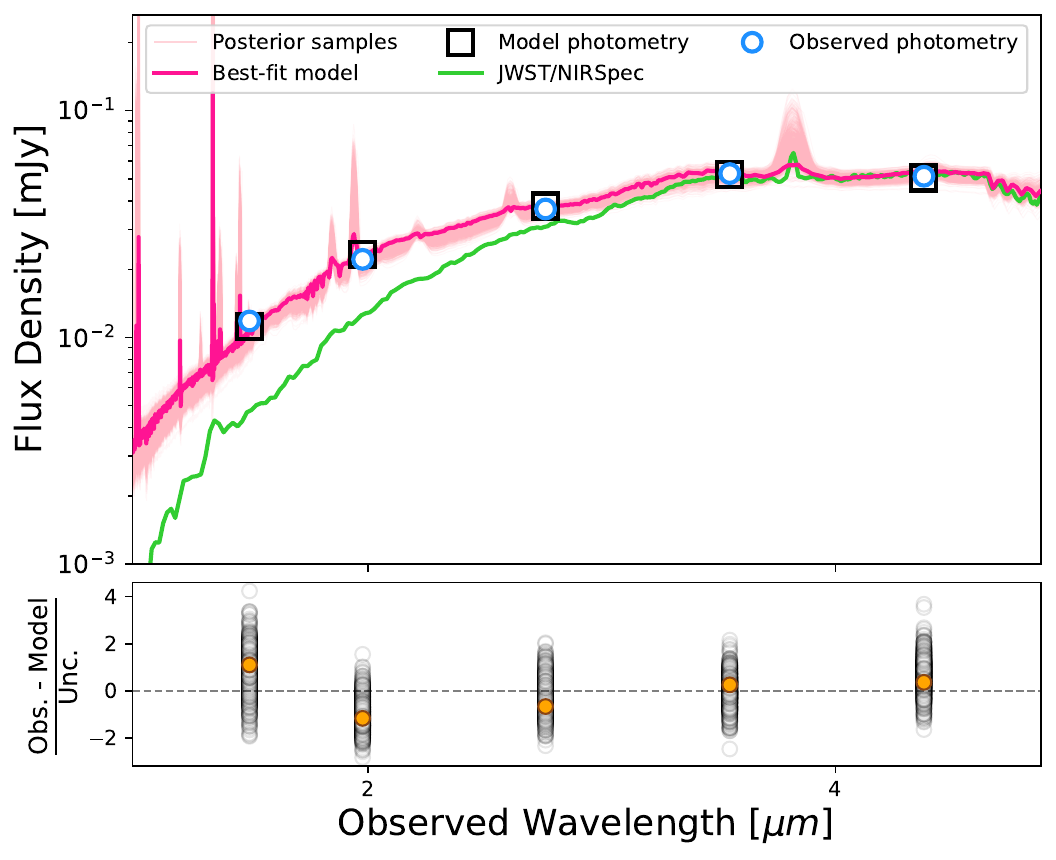}
\caption{\textbf{Top:} \texttt{Prospector} fit to the JWST host photometry, assuming a parametric star-formation history. The observations are shown in light blue, while the model photometry is shown as black squares. The full model spectral energy distribution is shown as a pink line. A sample of draws from the posterior are shown as light pink lines. We additionally plot in lime green the JWST/NIRSpec spectrum taken at the transient location, normalized to the F444W measurement \citep[originally presented in][]{Gompertz2025}.  The NIRSpec slit placement did not include the nucleus of the host galaxy, leading to the redder color of the spectrum than the host photometry. \textbf{Bottom:} The agreement between the best-fit model is shown in orange. The grey points show the agreement relative to a sample of draws from the posterior.} \label{fig:prospectorfit}
\end{figure}

Our model is defined as follows. We adopt a free redshift with a narrow, Gaussian prior with mean $z=1.036$ and width $0.004$ \citep[the redshift and uncertainty as measured from the JWST/NIRSpec spectrum of the source, ][see also Figure \ref{fig:prospectorfit}]{Gompertz2025}. Given the limited range of wavelengths covered by our data, and in particular a lack of ultraviolet and far-infrared constraints, we adopt the \cite{Calzetti2000TheGalaxies} attenuation law, which accurately models star-forming galaxies across a range of redshifts. We do not include dust emission due to the lack of rest-frame mid-infrared data. To break the age-metallicity degeneracy, we set the metallicity prior as a function of mass using the mass-metallicity relation from \cite{Gallazzi2005TheUniverse}. Following \cite{Leja2019AnSurvey}, we set the standard deviation of the prior to the 84th–-16th percentile range from \cite{Gallazzi2005TheUniverse}, or ${\sim}$twice the $z=0$ width, to account for redshift evolution. We include nebular emission, fixing the ionization parameter $\log U = -1$, which is higher than the typical assumed value $\log U = -2$ due to the more extreme ionizing fields observed in higher redshift galaxies \citep{Leja2019AnSurvey, Steidel2016RECONCILINGGALAXIES, Shapley:2015}. We fix the gas metallicity to the stellar metallicity.

We adopt a delayed-tau model star-formation history (SFH): $\textrm{SFR} \propto t e^{({-t/\tau})}$, where $t$ is measured relative to the galaxy age $t_{\rm age}$. We set the prior for $t_{\rm age}$ to be uniform up to the age of the universe at the galaxy redshift, $t_{\rm univ}$. We set the prior for $\tau$ to be log-uniform in the range 0.1--30\,Gyr. While a delayed-tau model is unlikely physically correct for this dataset, and non-parametric SFH models are expected to produce better results \citep[][]{Leja2019HowModels}, delayed-tau models have been used extensively in the literature and so enable like-to-like comparisons. We tested a seven bin, piecewise constant SFH with a StudentT prior with mean zero, width $0.3$, and two degrees-of-freedom \citep{Leja2019HowModels}, and do not find significant changes in mass, metallicity, dust content, or age.

We fiducially include only our JWST photometry in the fit, to ensure consistency of, e.g., adopted apertures. We also perform a fit including the HST photometry from \cite{Levan2025_BDE}. We choose to exclude the near-infrared photometry from \cite{Levan2025_BDE} and the $z$-band photometry from \cite{Carney2025}, because the extraction apertures are mismatched due to the significantly lower resolution of the observations.

We perform the fit using the \texttt{dynesty} dynamic nested sampler \citep[][]{Speagle2020DYNESTY:Evidences}. We use the following custom settings, as recommended by \cite{Leja2019AnSurvey}. We set \texttt{pfrac}$=1$ to prioritize sampling the posterior over computing the evidence. We use the \texttt{rslice} sampling method, $1500$ initial live points, initial stopping threshold $\Delta$log$\mathcal{Z}=0.01$, and $10^4$ target effective samples, with the aim of improving sampling of the highly generate posterior-space. We run until convergence. We fiducially present the parametric, JWST-only fit results, as previous efforts to model the host galaxy adopted parametric models and using only the JWST photometry ensures a consistent aperture for all observations.

The best-fit SED is shown in Figure~\ref{fig:prospectorfit}, and the corner plot is shown in the Appendix in Figure~\ref{fig:cornerplot}. We note a number of caveats with this fit. First, this galaxy has a strong dust lane, which causes it to violate the assumption of energy balance inherent to \texttt{Prospector}. The stellar library used (MILES) only supports metallicities $\log Z/Z_\odot < 0.19$, which is too low for the high-mass galaxy observed here. We also do not have sufficient data to overcome the mass-age-dust-metallicity degeneracy, although the mass-metallicity prior helps with this. The combination of the metallicity peaked near the prior edge and the strong correlations between parameters leads to poor posterior sampling, even with our improved \texttt{dynesty} convergence criteria. 

\begin{deluxetable}{lrr}
    \tablecaption{\texttt{Prospector} Best fit parameters.  Methods used to measure these values are in Sec. \ref{subsec:Proscpector}. \label{tbl:ProspectorTable}}
    \tablehead{\textbf{Parameter} & \textbf{Description} & \textbf{Value}}
    \startdata
        A$_{V}$ [mag] & Host Extinction & $2.8 \pm 0.3$ \\
        $\log(\tau)$ [Gyr] & Star-formation Timescale & $4.1^{+11.4}_{-3.4}$ \\
        $t_{age}/t_{univ}$ & Relative Age & $0.3^{+0.4}_{-0.2}$\\
        $\log(M_{*}/M_{\odot})$ & Stellar Mass & $11.0^{+0.2}_{-0.3}$ \\
        $\log(Z/Z_{\odot})$ & Metallicity & $-0.1^{+0.2}_{-0.3}$ \\
    \enddata
\end{deluxetable}

\section{Analysis}
\label{sec:analysis}

\subsection{Comparison Objects}
\label{subsec:compobj}

The intrinsic nature of GRB 250702B is still uncertain. However, the three leading hypotheses are: 1) an ultralong GRB; 2) a jetted TDE from a `massive' (or otherwise less massive, e.g., `intermediate-mass') black hole; and 3) a black hole merging with a helium star.  While there are several known ultralong GRBs and jetted TDEs, there have not yet been any confidently detected objects matching hypothesis 3. Hypothesis 3 does predict the emergence of a helium-rich SN, in contrast to the stripped envelope SNe observed with collapsar-GRBs \citep[e.g., GRB 111209A,][]{Kann2019}; however, our photometric data are not able to constrain various supernova types. For these reasons, we consider the feasibility of hypotheses 1 and 2.

\begin{figure}
\centering
\includegraphics[width = 1\columnwidth]{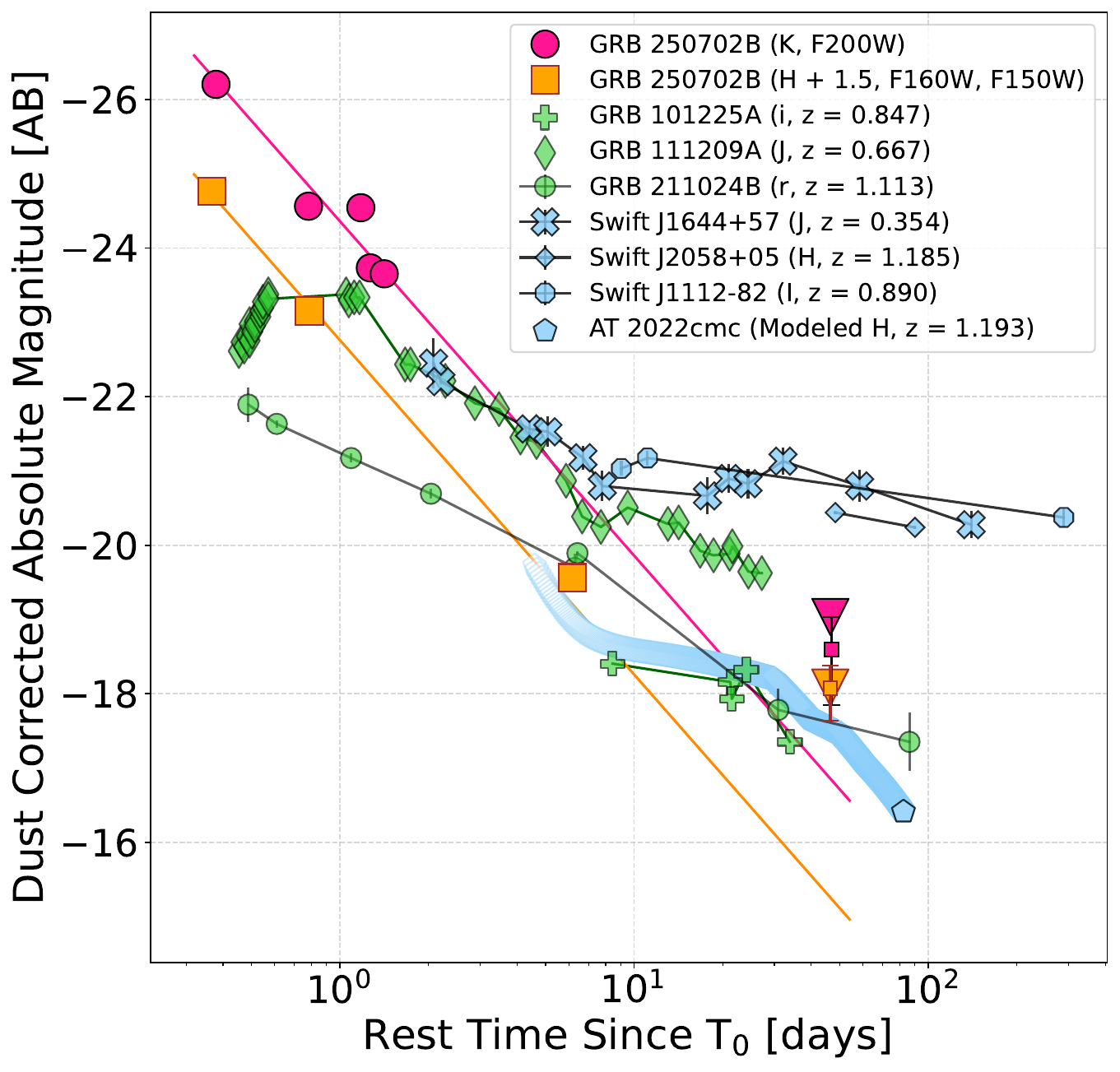}
\caption{Similar to Figure \ref{fig:lightcurve}, but now in dust-corrected absolute magnitude and rest-frame time and with no X-ray data and the jetted TDE (blue) and ULGRB (green) comparison objects. The observed-frame filter and redshift is listed for each object. These objects are described in Sec. \ref{subsec:compobj}. \label{fig:lightcurve_compobjs}}
\end{figure}

To investigate these scenarios, we select comparison objects, as detailed below and as shown in Figure \ref{fig:lightcurve_compobjs}. For the ultralong GRBs, we consider comparison objects of GRBs 101225A \citep[$z$ = 0.847;][]{Levan2014}, 111209A \citep[$z$ = 0.677;][]{Vreeswijk2011_grb111209Aredshift}, and 211024B \citep[$z$ = 1.113;][]{Fu2024_grb211024b}, with light curves drawn from \cite{Levan2014, Kann2018, Fu2024_grb211024b}, respectively. For the jetted TDEs, we select all four known objects: Swift J1644+57 \citep[$z$ = 0.354;][]{Zauderer2011}, Swift J2058+05 \citep[$z$ = 1.185;][]{Cenko2012}, Swift J1112-82 \citep[$z$ = 0.890;][]{Brown2015}, and AT 2022cmc \citep[$z$ = 1.193;][]{Andreoni2022}, with light curves drawn from \citet{Levan2016,Pasham2015, Brown2015, Hammerstein2026}, respectively. In the case of AT 2022cmc, we note that we use the extrapolated $H$ light curve from `Model 3' instead of an observed data set.  The comparisons are shown in Figure \ref{fig:lightcurve_compobjs}. We select these objects as they have sufficiently sampled optical/NIR light curves to comparably late times as our JWST observations of GRB 250702B.  We select light curves as close to rest-frame $r$ as possible so as to match the observed-frame $H$ and $K$ of GRB 250702B so as to minimize the uncertainty in the spectral shape and K-correction. To compare these objects to GRB 250702B, we extinction correct and convert to absolute magnitude with the assumption of a K-correction of $K_{corr} = 2.5 \times \log_{10}(1+z)$.

\subsection{ULGRB Interpretation}
\label{subsec:ulgrb}

\begin{figure*}
\centering
\includegraphics[width = 0.49\textwidth]{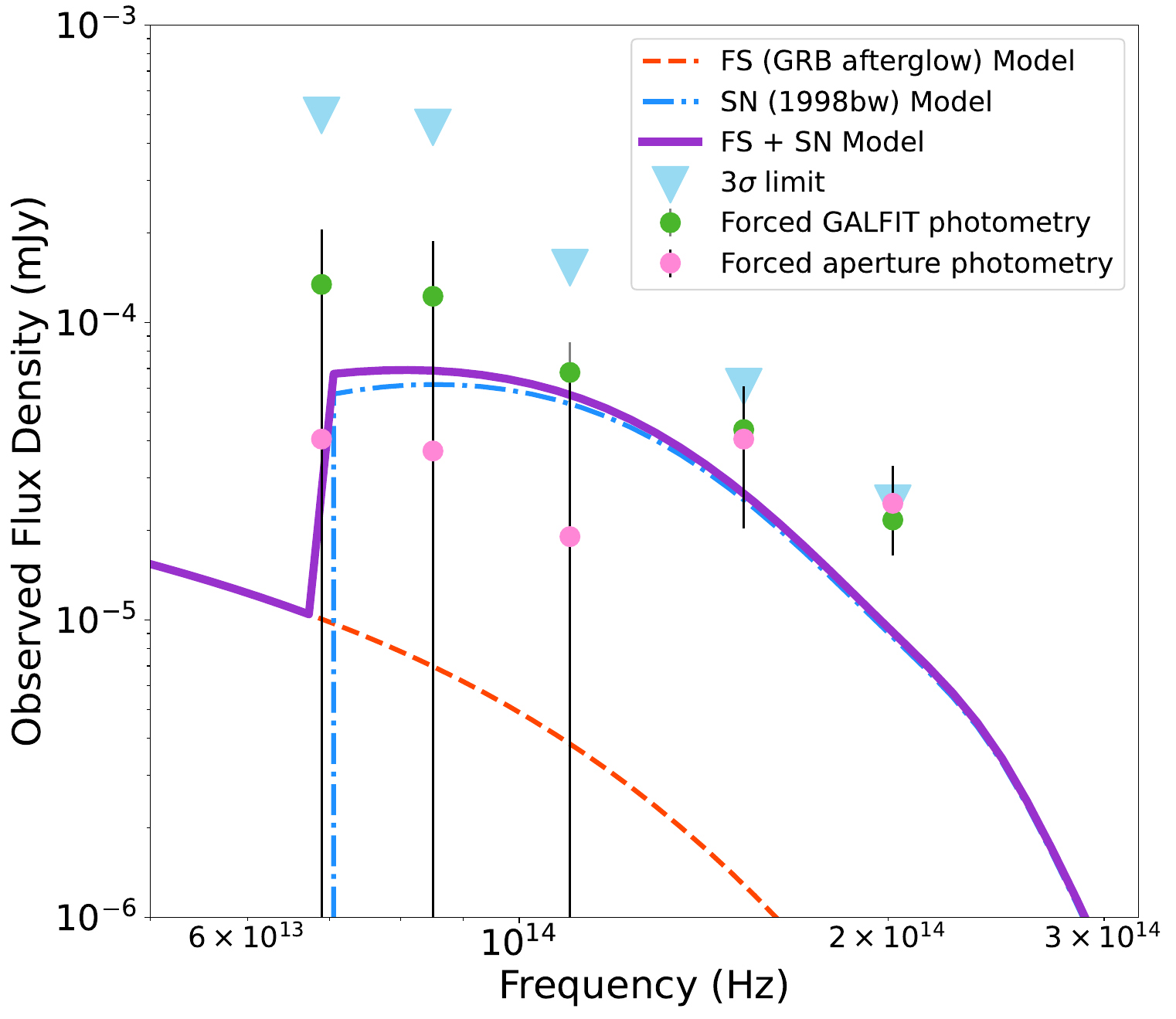}
\includegraphics[width = 0.49\textwidth]{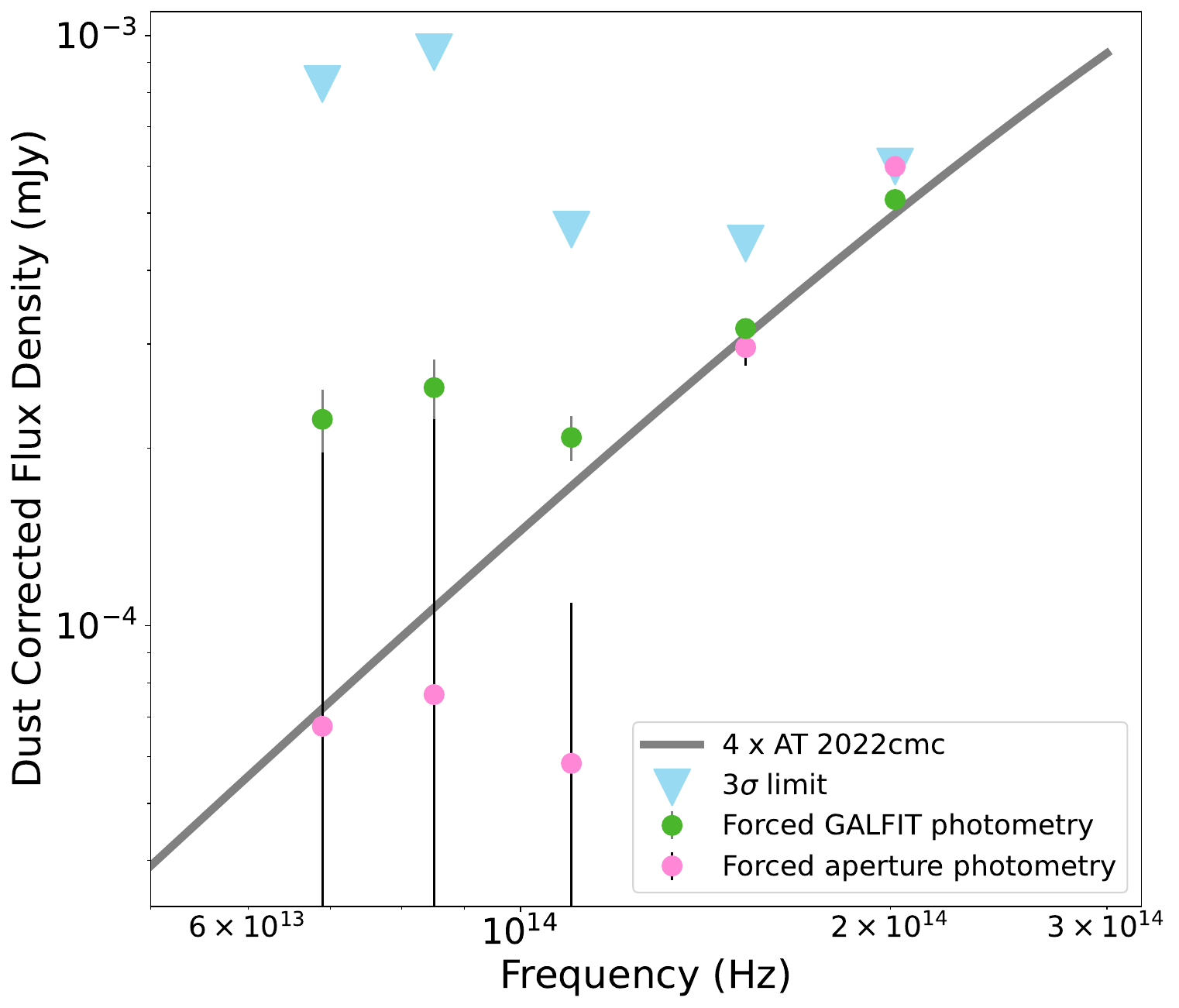}
\caption{\textbf{Left: } Observed transient upper limits (blue triangles) and forced photometry (pink and green triangles) compared to a forward shock model \citep[FS, dashed red,][]{Levan2025_BDE, Gompertz2025}, SN (1998bw, dot-dashed blue) model, and the sum of the two models (solid purple). The SN model used here does not extend beyond observer-frame $K$, which is the origin of the artificial sharp red cut off. The F150W$-$F200W color of the summed model matches the observed data well, however, the model is fainter than observed in GRB 250702B. \textbf{Right:} Extinction corrected photometry (same colors and symbols as in the left panel) compared to the 95 d (grey) SED model of the jetted TDE AT 2022cmc, from `Model 3' of \cite{Hammerstein2026}. The F150W$-$F200W color of the model matches the data of GRB 250702B, but the model is $\sim 4 \times$ fainter than the observed data. \label{fig:SN_fit}}
\end{figure*}

In this section, we first consider the implications of treating all flux in the forced photometry as light from the transient. A detection in F150W and F200W would clearly be in disagreement with assuming the X-ray light curve  slope ($t^{-1.8}$; \citealt{OConnor2025}) also applies the $H$ and $K$ decay (Figure \ref{fig:lightcurve_compobjs}). This behavior is inconsistent with that of a standard GRB afterglow model with the NIR and X-rays on the same synchrotron segment, where the light curve decays with the same power-law slope. Therefore, we investigate the possibility that the additional flux could be contributed from an accompanying supernova.  We test this hypothesis by combining the 95~d GRB afterglow model and the line-of-sight extinction curve from \cite{Gompertz2025, Levan2025_BDE} with a SN 1998bw  supernova model. As shown in the left panel of Figure \ref{fig:SN_fit}, this model under-predicts the observed photometry, although otherwise appears to match the shape of the transient SED.  To match the photometry, the accompanying supernova would need to be $\sim 2 \times$ as bright as SN 1998bw, seemingly in contradiction to the upper limits set by the spectroscopic non-detection in \cite{Gompertz2025}. We caution, however, that the SN luminosity is degenerate with the assumed host extinction, which is not well constrained \citep{Gompertz2025, Levan2025_BDE, Carney2025, OConnor2025}. 

We also present the $J$ light curve of a collapsar-driven ULGRB, GRB 111209A, in Figure \ref{fig:lightcurve_compobjs}, which shows late-time ($\Delta t>2$ d) similarity to the observations of GRB 250702B. GRB 111209A was accompanied by the very luminous SN 2011kl, which was modeled to be $\sim 1.6 \times$ L$_{\text{SN1998bw}}$ in the observed-frame $i$ and $z$ bands \citep{Kann2019}. The lack of well-cadenced, observed-frame $J$ data after the supernova peak prohibited a quantified luminosity multiplier in $J$, however \cite{Kann2019} qualify it as ``several times" brighter than SN 1998bw. The similarity of the light curves implies that GRB 250702B may have a similar intrinsic nature to GRB 111209A. The requirement of a supernova much brighter than SN 1998bw is also consistent with our modeling in Figure \ref{fig:SN_fit}. The requirement of a bright SN similar to SN 2011kl, however, is seemingly in contradiction to the non-detection of a SN in the earlier-phase JWST/NIRSpec spectrum at $\Delta t\sim 52$ d \citep{Gompertz2025}.

\subsection{Jetted TDE Interpretation}
\label{subsec:tde}

In an alternative progenitor hypothesis, all four known jetted TDEs show evidence for a flattening of the optical/NIR light curve at late times, as shown in Figure \ref{fig:lightcurve_compobjs}. Of note, the forced photometry for GRB 250702B is fainter than the plateaus of all jetted TDEs except for AT 2022cmc, though we caution this TDE light curve is a model extrapolation and not an observation \citep{Hammerstein2026}. Standard TDEs are observed to have late-time UV/optical plateaus, which are believed to be powered by the emergence of the accretion disk \citep{vanVelzen2019}, and, in this interpretation, \cite{Mummery2024} report a relation between the plateau luminosity and the black hole mass.  It is not clear if the plateau observed in jetted TDEs is powered by the same mechanism, and indeed, it also does not follow the relation found by \cite{Mummery2024} (i.e., when assuming the jetted TDE plateau magnitude, the calculated black hole mass is in conflict with other methods). We compare the predicted late-time color of AT 2022cmc to the observed color of GRB 250702B.  As shown in the right-panel of Figure \ref{fig:SN_fit}, the expected late-time F150W$-$F200W color of AT 2022cmc is in good agreement with the measured color of GRB 250702B from the forced photometry. The observed NIR plateau magnitude and late-time color of GRB 250702B are therefore broadly consistent with the jetted TDE.

On the other hand, if we treat the F150W/F200W observations as upper limits, the non-detections require any plateau phase to start at $\Delta t\gtrsim20 $ d, and at a dust-corrected plateau of $m_{H} \gtrsim 24.5 $ AB mag. This is a later onset time and a fainter plateau than that of the known jetted TDEs (save for the extrapolated model for AT 2022cmc) which could imply either a different mechanism powering the plateau or something different about the black hole at the center of the TDE (e.g., perhaps a different mass or a different spin).

\subsection{Other hypotheses}
\label{subsec:otherhypotheses}

Other possible explanations for the possible blue excess in the forced aperture photometry could be an optical light echo or, perhaps more simply, just underlying structure from the galaxy that is not well-subtracted.  The large uncertainty in the three reddest filters and the complicated residual make discerning explanations with any meaningful confidence difficult. More detailed multi-wavelength modeling of this source could better constrain the possible SN contribution and host extinction, though we leave this for future work.

\subsection{Host Comparison}
\label{subsec:HostComparison}

There have been no confidently classified off-nuclear jetted TDEs discovered to date, and as such, we are unable to compare the host observations of GRB 250702B to this subclass. While there have been at least four jetted TDEs discovered, all have been nuclear (or otherwise have had no observed host galaxy) and have properties that are well-established to correlate with host galaxy characteristics due to galaxy structure and kinematics. 

Off-nuclear TDEs (especially those that do not include a supermassive black hole) are not expected to have similarly correlated properties.  For these reasons, in considering the jetted TDE hypothesis, we restrict our comparison to the sample of off-nuclear TDE host galaxies. Galaxy mergers are expected to result in off-nuclear ``wandering" black holes and are, as such, predicted to be sites of off-nuclear TDEs \citep[e.g., ][]{Ricarte2021}. Of the classified off-nuclear TDEs, the central host galaxies of 3XMM J2150 \citep{Lin2018}, EP240222a \citep{Jin2025}, AT 2024tvd \citep{Yao2025}, TDE 2025abcr \citep{Stein2026}, and AT 2023mfm \citep{Li2026_offsetTDE} all have stellar masses of $M_{*} \sim 10^{10.9}$\,M$_{\odot}$.  Our \texttt{Prospector}-inferred stellar mass of $M_{*} \sim 10^{11.01^{+0.17}_{-0.24}}$\,M$_{\odot}$ is consistent with the masses of the offset TDE host galaxy sample. GRB 250702B appears to be in the disk of the host galaxy, and we see no evidence for a companion or dwarf galaxy at the transient location. This may favor an interpretation of an intermediate-mass black hole formed in the disk over a wandering SMBH in the TDE interpretation.

To investigate the feasibility of the interpretation of GRB 250702B as a true GRB, we compare measured host galaxy properties to the broader long GRB host galaxy sample.  The largest and most complete samples are those presented in \cite{Blanchard2016} and SHOALS \citep{SHOALSI, SHOALSII}.  The majority of the HST observations of GRB host galaxies presented in \cite{Blanchard2016} were performed at bluer rest-frame wavelengths than for GRB 250702B. To select the most comparable subsample, we therefore limit our comparison to those host galaxies at similar redshifts that were observed in filters with rest wavelengths equivalent to our F150W observations (roughly, rest-frame $r$ band). This results in a subsample of 9 GRB hosts with a redshift range of $z \sim$ 0.8--1.3.  

For these 9 GRB host galaxies, we collect measurements of their fractional flux, GRB offset, apparent magnitude, and galaxy half-light radius from the tables presented in \cite{Blanchard2016}.  To measure the half-light radius of the host galaxy of GRB 250702B in F150W, we use \texttt{Source Extractor} \citep{Bertin1996}.  We then also use the center of the detected ellipse from \texttt{Source Extractor} as the center of the galaxy and measure an offset from this position to that of the transient.  To normalize this offset measurement, we divide by the half-light radius.  Finally, to measure the fractional flux, we first subtract the diffraction-spike-subtracted image by the median off-galaxy sky pixel. We then use the same aperture as was used for the host aperture photometry and select all pixels with flux values at least 1$\sigma$ above the sky. After this flux cut, we calculate the ratio of the sum of all the pixels with flux equal-to-or-less-than the flux of the transient pixel divided by the sum of the flux of all of the pixels.  We report a fractional flux in F150W of 0.730, a half-light radius of 0.423\arcsec, and a host normalized offset of 1.59. We also compute absolute magnitudes using the F150W  magnitude for GRB 250702B in comparison to the sample magnitudes in the observed filters closest to rest-frame $r$.  

The SHOALS \citep{SHOALSI, SHOALSII} sample presents observed-frame $3.6\ \mu m$ observations of GRB host galaxies. To select an appropriate subsample for comparison, we also implement a redshift cut of $0.8 < z < 1.3$, to be consistent with the previous analysis.  From this cut, we have 15 GRB host galaxies for which we collect host masses and compute absolute magnitudes using the observed-frame $3.6\ \mu m$ photometry.  Of these, five host galaxies have only upper limits in mass and lower limits in magnitude.  For GRB 250702B, we use the host mass we infer from our \texttt{Prospector} modeling and the measured F356W host magnitude.  We again compute absolute magnitudes using the F356W magnitude for GRB 250702B in comparison to the sample magnitudes in observer-frame $3.6 \mu m$, which is roughly rest-frame $H$.

\begin{figure*}
\centering
\includegraphics[width = \textwidth]{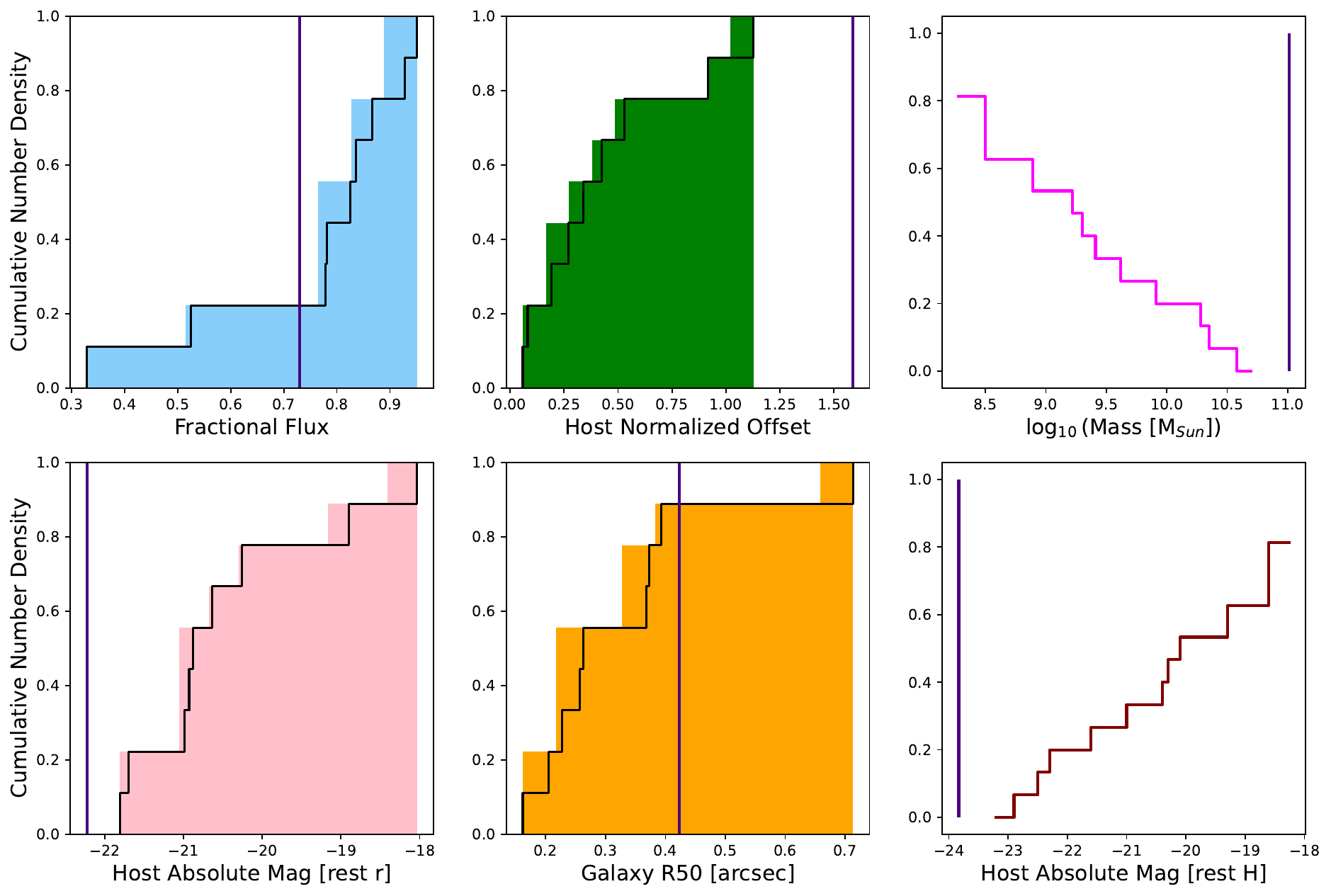}
\caption{Cumulative histograms (shaded regions) and eCDFs (solid lines) for a subsample of GRB host galaxies at $0.8 < z < 1.3$.  In clockwise order starting at the top left are the fractional flux, host normalized offset, stellar mass, host absolute magnitude in observed-frame $3.6\ \mu m$ ($\approx$ rest-frame $H$), half-light radius in arcseconds, and the host absolute magnitude in the filter closest to rest-frame $r$. For both absolute magnitudes, only a $K$-correction of $-2.5\log(1+z)$ has been applied. The value of each parameter for the host galaxy of GRB 250702B is marked with an indigo vertical line. Further details on these comparisons are described in Section \ref{subsec:HostComparison}} \label{fig:compplots}
\end{figure*}

We present cumulative histograms and the empirical cumulative distribution functions (eCDFs) for the properties of the comparison GRB host galaxies in Figure \ref{fig:compplots}. We note that since the SHOALS subsample includes censored data (i.e., upper limits), we can only present eCDFs.

It is clear that the host of GRB 250702B is an outlier in rest-frame $r$ host magnitude (brightest), rest-frame $H$ host magnitude (brightest), and stellar mass (most massive). It is notable that this host galaxy is still brighter than the other members of the subsample despite the presence of a prominent dust lane absorbing flux. Similar conclusions about the host galaxy were reported in previous studies \citep{Levan2025_BDE, Carney2025, Gompertz2025}. In this small redshift-matched subsample of 9 galaxies, GRB 250702B is also the most offset from its host. However, this is an artifact of the small comparison sample, as \cite{Blanchard2016} find that across all redshifts, $\sim$15\% of long GRBs have higher normalized offsets. The host is otherwise consistent with the distributions of fractional flux and half-light radius for the other GRB hosts.

\section{Conclusions}
\label{sec:conclusion}

We present new JWST/NIRCam imaging of the field of GRB 250702B in the  F150W, F200W, F277W, F356W, and F444W filters.  Our observations reveal a single host galaxy with a prominent dust lane observed nearly fully edge-on.  Our observations also reveal a complicated environment near the location of the transient.  We report possible $\sim 3 \sigma$ detections of GRB 250702B in F150W and F200W.  

If these detections are real, these data are consistent with the color and late-time temporal behavior (i.e., a plateau) of jetted TDEs. However, the plateau of GRB 250702B is $\sim$2 mag fainter than 3 of the 4 known jetted TDEs.  This fainter plateau is broadly consistent with expectations of a jetted TDE from a black hole with a mass $ M_{BH} \leq 10^5$~M$_{\odot}$. If GRB 250702B were produced by a jetted TDE, it would represent the first jetted TDE to be associated with a black hole less massive than a central SMBH. It would also represent the first off-nuclear jetted TDE. There are only four known jetted TDEs and five robustly identified off-nuclear TDEs, which makes it difficult to make strong conclusions as to the diversity of the physical origins of the observational differences between jetted, off-nuclear, and different-mass TDEs.

These data are also consistent with the extrapolated GRB afterglow model \citep{Gompertz2025, Levan2025_BDE} with the addition of a supernova component, assuming a spectrum like SN 1998bw at twice the luminosity.  This afterglow+SN model is in possible contradiction with the SN limits placed by the JWST/NIRSpec spectrum \citep{Gompertz2025}. However, the host extinction is poorly constrained and is degenerate with the assumed SN luminosity. When treating the F150W and F200W measurements as non-detections, these limits are consistent with, but do not further constrain, the extrapolated afterglow temporal decay of $t^{-1.8}$ as measured from X-ray observations \citep{OConnor2025}. 

If GRB 250702B is a collapsar-driven ULGRB with a supernova luminosity at $2\times$ SN 1998bw (or even fainter than this), GRB 250702B would represent just the second ULGRB with an associated SN.  The other SN-ULGRB was the extremely luminous SN 2011kl, associated with GRB 111209A, which showed transitional features between SNe Ic-BL and SLSNe \citep{Greiner2015, Kann2019}.  The diversity between these ULGRB-associated SNe (perhaps even especially in the lack of identified SNe with other ULGRBs), while somewhat similar to that seen with collapsar-driven LGRBs, invites the hypothesis that the ultralong GRB class may not be homogeneous in progenitor path. Further along this interpretation, the early X-ray emission observed 24 hrs prior with the Einstein Probe is unprecedented among GRBs (including ultralong GRBs) and seemingly incompatible with theory on the jet originating during the collapse of the star.  It is possible, however, that this X-ray emission, especially the soft X-ray emission, was simply missed for these other ultralong GRBs. Further, we have observed ``precursor" X-ray emission with several long GRBs \citep[e.g.,][]{Murakami1991, Lazzati2005, Guidorzi2025}, though none as early as 24 hr prior as in GRB 250702B.  

For future study of GRB 250702B, there is clear need for a panchromatic study to resolve some of the still open questions.  In particular, we do not re-fit a GRB afterglow + SN model to our data and instead simply apply the extrapolated model from \citet{Gompertz2025} and \citet{Levan2025_BDE} and fit for a supernova luminosity.  This particular modeling would be best completed with a full multi-wavelength data set, including late-time radio data, so as to better constrain any deviation from the late-time slope measured from the X-ray emission.  Finally, due to the complicated structure of the host galaxy of GRB 250702B and the existence of the strong dust lane, there is ambiguity as to whether there is a transient detection in F150W and F200W.  This detected emission could simply be unrelated coincident emission, given the complex structure of the galaxy. Template imaging taken with JWST would significantly help to resolve this ambiguity.  GRB 250702B is an extreme, enigmatic transient that challenges many of our theories on collapsar GRBs and the origins of TDEs. As such, we encourage further study of this object.

\begin{acknowledgments}
We thank Federica Bianco, Charlotte Olsen, and Jay Strader for useful discussion and suggestions on plot aesthetics. We thank Jennifer Lotz and the team at STScI for approval of this DD request, and we thank Alison Vick and Ben Sunnquist for their assistance in preparing the JWST observations. This work is based on observations made with the NASA/ESA/CSA JWST. The data were obtained from the Mikulski Archive for Space Telescopes at the Space Telescope Science Institute, which is operated by the Association of Universities for Research in Astronomy, Inc., under NASA contract NAS 5-03127 for JWST. These observations are associated with JWST program ID: DD 9447. Support for this program at Rutgers University was provided by NASA through grant JWST-GO-09447.002-A. 

Based on observations made with the Nordic Optical Telescope, owned in collaboration by the University of Turku and Aarhus University, and operated jointly by Aarhus University, the University of Turku and the University of Oslo, representing Denmark, Finland and Norway, the University of Iceland and Stockholm University at the Observatorio del Roque de los Muchachos, La Palma, Spain, of the Instituto de Astrofisica de Canarias. The NOT data were obtained under program ID 71-203.

H.S. acknowledges partial salary support from a Moore Foundation Postdoctoral Fellowship Grant to Rutgers University. S.W.J. gratefully acknowledges support from a Guggenheim Fellowship.

B.O. is supported by the McWilliams Postdoctoral Fellowship in the McWilliams Center for Cosmology and Astrophysics at Carnegie Mellon University. 

L.G. acknowledges financial support from CSIC, MCIN and AEI 10.13039/501100011033 under projects PID2023-151307NB-I00, PIE 20215AT016, CEX2020-001058-M, and by the MaX-CSIC Excellence Award MaX4-SOMMA-ICE. This research has made use of the SVO Filter Profile Service ``Carlos Rodrigo," \citep{SVO_cite1, SVO_cite2, SVO_cite3} funded by MCIN/AEI/10.13039/501100011033/ through grant PID2023-146210NB-I00.  This research has made use of the WebPlotDigitizer \citep{WebPlotDigitizer}.

Portions of the spectral energy distribution fitting code were developed with assistance from Claude. All code was reviewed, tested, and validated by the authors.

The VISTA Hemisphere Survey data products served at Astro Data Lab are based on observations collected at the European Organisation for Astronomical Research in the Southern Hemisphere under ESO programme 179.A-2010, and/or data products created thereof.

This research has made use of the NASA/IPAC Extragalactic Database, which is funded by the National Aeronautics and Space Administration and operated by the California Institute of Technology.

This research made use of Photutils, an Astropy package for
detection and photometry of astronomical sources \citep{larry_bradley_2025}.
\end{acknowledgments}

\facilities{Chandra, HST, JWST, NOT, NuSTAR, Swift (XRT)}

\bibliography{myrefs}

@ARTICLE{Peng2010,
       author = {{Peng}, Chien Y. and {Ho}, Luis C. and {Impey}, Chris D. and {Rix}, Hans-Walter},
        title = "{Detailed Decomposition of Galaxy Images. II. Beyond Axisymmetric Models}",
      journal = {\aj},
         year = 2010,
        month = jun,
       volume = {139},
       number = {6},
        pages = {2097-2129},
          doi = {10.1088/0004-6256/139/6/2097},
archivePrefix = {arXiv},
       eprint = {0912.0731},
 primaryClass = {astro-ph.CO},
       adsurl = {https://ui.adsabs.harvard.edu/abs/2010AJ....139.2097P}
}

@ARTICLE{Levan2011,
       author = {{Levan}, A.~J. and {Tanvir}, N.~R. and {Cenko}, S.~B. and {Perley}, D.~A. and {Wiersema}, K. and {Bloom}, J.~S. and {Fruchter}, A.~S. and {de Ugarte Postigo}, A. and {O'Brien}, P.~T. and {Butler}, N. and {van der Horst}, A.~J. and {Leloudas}, G. and {Morgan}, A.~N. and {Misra}, K. and {Bower}, G.~C. and {Farihi}, J. and {Tunnicliffe}, R.~L. and {Modjaz}, M. and {Silverman}, J.~M. and {Hjorth}, J. and {Th{\"o}ne}, C. and {Cucchiara}, A. and {Cer{\'o}n}, J.~M. Castro and {Castro-Tirado}, A.~J. and {Arnold}, J.~A. and {Bremer}, M. and {Brodie}, J.~P. and {Carroll}, T. and {Cooper}, M.~C. and {Curran}, P.~A. and {Cutri}, R.~M. and {Ehle}, J. and {Forbes}, D. and {Fynbo}, J. and {Gorosabel}, J. and {Graham}, J. and {Hoffman}, D.~I. and {Guziy}, S. and {Jakobsson}, P. and {Kamble}, A. and {Kerr}, T. and {Kasliwal}, M.~M. and {Kouveliotou}, C. and {Kocevski}, D. and {Law}, N.~M. and {Nugent}, P.~E. and {Ofek}, E.~O. and {Poznanski}, D. and {Quimby}, R.~M. and {Rol}, E. and {Romanowsky}, A.~J. and {S{\'a}nchez-Ram{\'\i}rez}, R. and {Schulze}, S. and {Singh}, N. and {van Spaandonk}, L. and {Starling}, R.~L.~C. and {Strom}, R.~G. and {Tello}, J.~C. and {Vaduvescu}, O. and {Wheatley}, P.~J. and {Wijers}, R.~A.~M.~J. and {Winters}, J.~M. and {Xu}, D.},
        title = "{An Extremely Luminous Panchromatic Outburst from the Nucleus of a Distant Galaxy}",
      journal = {Science},
         year = 2011,
        month = jul,
       volume = {333},
       number = {6039},
        pages = {199},
          doi = {10.1126/science.1207143},
archivePrefix = {arXiv},
       eprint = {1104.3356},
 primaryClass = {astro-ph.HE},
       adsurl = {https://ui.adsabs.harvard.edu/abs/2011Sci...333..199L}
}

@ARTICLE{Li2026_offsetTDE,
       author = {{Li}, Wenkai and {Christy}, Collin T. and {Alexander}, Kate D. and {Sfaradi}, Itai and {Huang}, Xinya and {Jiang}, Ning and {Laskar}, Tanmoy and {Mummery}, Andrew and {Franz}, Noah and {Goodwin}, Adelle J. and {Golay}, Walter W. and {Margutti}, Raffaella and {Chornock}, Ryan and {Zhu}, Jiazheng and {van Velzen}, Sjoert and {Cendes}, Yvette and {Lu}, Wenbin and {Lynch}, Jimmy},
        title = "{VLA Observations Confirm AT 2023mfm as an Off-nuclear Tidal Disruption Event}",
      journal = {arXiv e-prints},
         year = 2026,
        month = jun,
          eid = {arXiv:2606.06595},
        pages = {arXiv:2606.06595},
archivePrefix = {arXiv},
       eprint = {2606.06595},
 primaryClass = {astro-ph.HE},
       adsurl = {https://ui.adsabs.harvard.edu/abs/2026arXiv260606595L}
}

@ARTICLE{Starling2011,
       author = {{Starling}, R.~L.~C. and {Wiersema}, K. and {Levan}, A.~J. and {Sakamoto}, T. and {Bersier}, D. and {Goldoni}, P. and {Oates}, S.~R. and {Rowlinson}, A. and {Campana}, S. and {Sollerman}, J. and {Tanvir}, N.~R. and {Malesani}, D. and {Fynbo}, J.~P.~U. and {Covino}, S. and {D'Avanzo}, P. and {O'Brien}, P.~T. and {Page}, K.~L. and {Osborne}, J.~P. and {Vergani}, S.~D. and {Barthelmy}, S. and {Burrows}, D.~N. and {Cano}, Z. and {Curran}, P.~A. and {de Pasquale}, M. and {D'Elia}, V. and {Evans}, P.~A. and {Flores}, H. and {Fruchter}, A.~S. and {Garnavich}, P. and {Gehrels}, N. and {Gorosabel}, J. and {Hjorth}, J. and {Holland}, S.~T. and {van der Horst}, A.~J. and {Hurkett}, C.~P. and {Jakobsson}, P. and {Kamble}, A.~P. and {Kouveliotou}, C. and {Kuin}, N.~P.~M. and {Kaper}, L. and {Mazzali}, P.~A. and {Nugent}, P.~E. and {Pian}, E. and {Stamatikos}, M. and {Th{\"o}ne}, C.~C. and {Woosley}, S.~E.},
        title = "{Discovery of the nearby long, soft GRB 100316D with an associated supernova}",
      journal = {\mnras},
         year = 2011,
        month = mar,
       volume = {411},
       number = {4},
        pages = {2792-2803},
          doi = {10.1111/j.1365-2966.2010.17879.x},
archivePrefix = {arXiv},
       eprint = {1004.2919},
 primaryClass = {astro-ph.CO},
       adsurl = {https://ui.adsabs.harvard.edu/abs/2011MNRAS.411.2792S}
}

@ARTICLE{Cano2011,
       author = {{Cano}, Z. and {Bersier}, D. and {Guidorzi}, C. and {Kobayashi}, S. and {Levan}, A.~J. and {Tanvir}, N.~R. and {Wiersema}, K. and {D'Avanzo}, P. and {Fruchter}, A.~S. and {Garnavich}, P. and {Gomboc}, A. and {Gorosabel}, J. and {Kasen}, D. and {Kopa{\v{c}}}, D. and {Margutti}, R. and {Mazzali}, P.~A. and {Melandri}, A. and {Mundell}, C.~G. and {Nugent}, P.~E. and {Pian}, E. and {Smith}, R.~J. and {Steele}, I. and {Wijers}, R.~A.~M.~J. and {Woosley}, S.~E.},
        title = "{XRF 100316D/SN 2010bh and the Nature of Gamma-Ray Burst Supernovae}",
      journal = {\apj},
         year = 2011,
        month = oct,
       volume = {740},
       number = {1},
          eid = {41},
        pages = {41},
          doi = {10.1088/0004-637X/740/1/41},
archivePrefix = {arXiv},
       eprint = {1104.5141},
 primaryClass = {astro-ph.SR},
       adsurl = {https://ui.adsabs.harvard.edu/abs/2011ApJ...740...41C}
}

@ARTICLE{Olivares2012,
       author = {{Olivares E.}, F. and {Greiner}, J. and {Schady}, P. and {Rau}, A. and {Klose}, S. and {Kr{\"u}hler}, T. and {Afonso}, P.~M.~J. and {Updike}, A.~C. and {Nardini}, M. and {Filgas}, R. and {Nicuesa Guelbenzu}, A. and {Clemens}, C. and {Elliott}, J. and {Kann}, D.~A. and {Rossi}, A. and {Sudilovsky}, V.},
        title = "{The fast evolution of SN 2010bh associated with XRF 100316D}",
      journal = {\aap},
         year = 2012,
        month = mar,
       volume = {539},
          eid = {A76},
        pages = {A76},
          doi = {10.1051/0004-6361/201117929},
archivePrefix = {arXiv},
       eprint = {1110.4109},
 primaryClass = {astro-ph.HE},
       adsurl = {https://ui.adsabs.harvard.edu/abs/2012A&A...539A..76O}
}

@ARTICLE{Bufano2012,
       author = {{Bufano}, Filomena and {Pian}, Elena and {Sollerman}, Jesper and {Benetti}, Stefano and {Pignata}, Giuliano and {Valenti}, Stefano and {Covino}, Stefano and {D'Avanzo}, Paolo and {Malesani}, Daniele and {Cappellaro}, Enrico and {Della Valle}, Massimo and {Fynbo}, Johan and {Hjorth}, Jens and {Mazzali}, Paolo A. and {Reichart}, Daniel E. and {Starling}, Rhaana L.~C. and {Turatto}, Massimo and {Vergani}, Susanna D. and {Wiersema}, Klass and {Amati}, Lorenzo and {Bersier}, David and {Campana}, Sergio and {Cano}, Zach and {Castro-Tirado}, Alberto J. and {Chincarini}, Guido and {D'Elia}, Valerio and {de Ugarte Postigo}, Antonio and {Deng}, Jinsong and {Ferrero}, Patrizia and {Filippenko}, Alexei V. and {Goldoni}, Paolo and {Gorosabel}, Javier and {Greiner}, Jochen and {Hammer}, Francois and {Jakobsson}, Pall and {Kaper}, Lex and {Kawabata}, Koji S. and {Klose}, Sylvio and {Levan}, Andrew J. and {Maeda}, Keiichi and {Masetti}, Nicola and {Milvang-Jensen}, Bo and {Mirabel}, Felix I. and {M{\o}ller}, Palle and {Nomoto}, Ken'ichi and {Palazzi}, Eliana and {Piranomonte}, Silvia and {Salvaterra}, Ruben and {Stratta}, Giulia and {Tagliaferri}, Gianpiero and {Tanaka}, Masaomi and {Tanvir}, Nial R. and {Wijers}, Ralph A.~M.~J.},
        title = "{The Highly Energetic Expansion of SN 2010bh Associated with GRB 100316D}",
      journal = {\apj},
         year = 2012,
        month = jul,
       volume = {753},
       number = {1},
          eid = {67},
        pages = {67},
          doi = {10.1088/0004-637X/753/1/67},
archivePrefix = {arXiv},
       eprint = {1111.4527},
 primaryClass = {astro-ph.HE},
       adsurl = {https://ui.adsabs.harvard.edu/abs/2012ApJ...753...67B}
}

@ARTICLE{Bufano2011,
       author = {{Bufano}, F. and {Benetti}, S. and {Sollerman}, J. and {Pian}, E. and {Cupani}, G.},
        title = "{Studying the SN-GRB connection with X-shooter: The GRB 100316D / SN 2010bh case}",
      journal = {Astronomische Nachrichten},
         year = 2011,
        month = mar,
       volume = {332},
       number = {3},
        pages = {262},
          doi = {10.1002/asna.201111531},
archivePrefix = {arXiv},
       eprint = {1103.5298},
 primaryClass = {astro-ph.CO},
       adsurl = {https://ui.adsabs.harvard.edu/abs/2011AN....332..262B}
}

@ARTICLE{Wang2006,
       author = {{Wang}, Xiang-Yu and {M{\'e}sz{\'a}ros}, Peter},
        title = "{GeV Photons from the Upscattering of Supernova Shock Breakout X-Rays by an Outside Gamma-Ray Burst Jet}",
      journal = {\apjl},
         year = 2006,
        month = jun,
       volume = {643},
       number = {2},
        pages = {L95-L98},
          doi = {10.1086/505142},
archivePrefix = {arXiv},
       eprint = {astro-ph/0603719},
 primaryClass = {astro-ph},
       adsurl = {https://ui.adsabs.harvard.edu/abs/2006ApJ...643L..95W}
}

@ARTICLE{Modjaz2006,
       author = {{Modjaz}, M. and {Stanek}, K.~Z. and {Garnavich}, P.~M. and {Berlind}, P. and {Blondin}, S. and {Brown}, W. and {Calkins}, M. and {Challis}, P. and {Diamond-Stanic}, A.~M. and {Hao}, H. and {Hicken}, M. and {Kirshner}, R.~P. and {Prieto}, J.~L.},
        title = "{Early-Time Photometry and Spectroscopy of the Fast Evolving SN 2006aj Associated with GRB 060218}",
      journal = {\apjl},
         year = 2006,
        month = jul,
       volume = {645},
       number = {1},
        pages = {L21-L24},
          doi = {10.1086/505906},
archivePrefix = {arXiv},
       eprint = {astro-ph/0603377},
 primaryClass = {astro-ph},
       adsurl = {https://ui.adsabs.harvard.edu/abs/2006ApJ...645L..21M}
}

@ARTICLE{Cobb2006,
       author = {{Cobb}, B.~E. and {Bailyn}, C.~D. and {van Dokkum}, P.~G. and {Natarajan}, P.},
        title = "{SN 2006aj and the Nature of Low-Luminosity Gamma-Ray Bursts}",
      journal = {\apjl},
         year = 2006,
        month = jul,
       volume = {645},
       number = {2},
        pages = {L113-L116},
          doi = {10.1086/506271},
archivePrefix = {arXiv},
       eprint = {astro-ph/0603832},
 primaryClass = {astro-ph},
       adsurl = {https://ui.adsabs.harvard.edu/abs/2006ApJ...645L.113C}
}

@ARTICLE{Sollerman2006,
       author = {{Sollerman}, J. and {Jaunsen}, A.~O. and {Fynbo}, J.~P.~U. and {Hjorth}, J. and {Jakobsson}, P. and {Stritzinger}, M. and {F{\'e}ron}, C. and {Laursen}, P. and {Ovaldsen}, J.-E. and {Selj}, J. and {Th{\"o}ne}, C.~C. and {Xu}, D. and {Davis}, T. and {Gorosabel}, J. and {Watson}, D. and {Duro}, R. and {Ilyin}, I. and {Jensen}, B.~L. and {Lysfjord}, N. and {Marquart}, T. and {Nielsen}, T.~B. and {N{\"a}r{\"a}nen}, J. and {Schwarz}, H.~E. and {Walch}, S. and {Wold}, M. and {{\"O}stlin}, G.},
        title = "{Supernova 2006aj and the associated X-Ray Flash 060218}",
      journal = {\aap},
         year = 2006,
        month = aug,
       volume = {454},
       number = {2},
        pages = {503-509},
          doi = {10.1051/0004-6361:20065226},
archivePrefix = {arXiv},
       eprint = {astro-ph/0603495},
 primaryClass = {astro-ph},
       adsurl = {https://ui.adsabs.harvard.edu/abs/2006A&A...454..503S}
}

@ARTICLE{Campana2006,
       author = {{Campana}, S. and {Mangano}, V. and {Blustin}, A.~J. and {Brown}, P. and {Burrows}, D.~N. and {Chincarini}, G. and {Cummings}, J.~R. and {Cusumano}, G. and {Della Valle}, M. and {Malesani}, D. and {M{\'e}sz{\'a}ros}, P. and {Nousek}, J.~A. and {Page}, M. and {Sakamoto}, T. and {Waxman}, E. and {Zhang}, B. and {Dai}, Z.~G. and {Gehrels}, N. and {Immler}, S. and {Marshall}, F.~E. and {Mason}, K.~O. and {Moretti}, A. and {O'Brien}, P.~T. and {Osborne}, J.~P. and {Page}, K.~L. and {Romano}, P. and {Roming}, P.~W.~A. and {Tagliaferri}, G. and {Cominsky}, L.~R. and {Giommi}, P. and {Godet}, O. and {Kennea}, J.~A. and {Krimm}, H. and {Angelini}, L. and {Barthelmy}, S.~D. and {Boyd}, P.~T. and {Palmer}, D.~M. and {Wells}, A.~A. and {White}, N.~E.},
        title = "{The association of GRB 060218 with a supernova and the evolution of the shock wave}",
      journal = {\nat},
         year = 2006,
        month = aug,
       volume = {442},
       number = {7106},
        pages = {1008-1010},
          doi = {10.1038/nature04892},
archivePrefix = {arXiv},
       eprint = {astro-ph/0603279},
 primaryClass = {astro-ph},
       adsurl = {https://ui.adsabs.harvard.edu/abs/2006Natur.442.1008C}
}

@ARTICLE{Pian2006,
       author = {{Pian}, E. and {Mazzali}, P.~A. and {Masetti}, N. and {Ferrero}, P. and {Klose}, S. and {Palazzi}, E. and {Ramirez-Ruiz}, E. and {Woosley}, S.~E. and {Kouveliotou}, C. and {Deng}, J. and {Filippenko}, A.~V. and {Foley}, R.~J. and {Fynbo}, J.~P.~U. and {Kann}, D.~A. and {Li}, W. and {Hjorth}, J. and {Nomoto}, K. and {Patat}, F. and {Sauer}, D.~N. and {Sollerman}, J. and {Vreeswijk}, P.~M. and {Guenther}, E.~W. and {Levan}, A. and {O'Brien}, P. and {Tanvir}, N.~R. and {Wijers}, R.~A.~M.~J. and {Dumas}, C. and {Hainaut}, O. and {Wong}, D.~S. and {Baade}, D. and {Wang}, L. and {Amati}, L. and {Cappellaro}, E. and {Castro-Tirado}, A.~J. and {Ellison}, S. and {Frontera}, F. and {Fruchter}, A.~S. and {Greiner}, J. and {Kawabata}, K. and {Ledoux}, C. and {Maeda}, K. and {M{\o}ller}, P. and {Nicastro}, L. and {Rol}, E. and {Starling}, R.},
        title = "{An optical supernova associated with the X-ray flash XRF 060218}",
      journal = {\nat},
         year = 2006,
        month = aug,
       volume = {442},
       number = {7106},
        pages = {1011-1013},
          doi = {10.1038/nature05082},
archivePrefix = {arXiv},
       eprint = {astro-ph/0603530},
 primaryClass = {astro-ph},
       adsurl = {https://ui.adsabs.harvard.edu/abs/2006Natur.442.1011P}
}

@ARTICLE{Burns2023,
       author = {{Burns}, Eric and {Svinkin}, Dmitry and {Fenimore}, Edward and {Kann}, D. Alexander and {Ag{\"u}{\'\i} Fern{\'a}ndez}, Jos{\'e} Feliciano and {Frederiks}, Dmitry and {Hamburg}, Rachel and {Lesage}, Stephen and {Temiraev}, Yuri and {Tsvetkova}, Anastasia and {Bissaldi}, Elisabetta and {Briggs}, Michael S. and {Dalessi}, Sarah and {Dunwoody}, Rachel and {Fletcher}, Cori and {Goldstein}, Adam and {Hui}, C. Michelle and {Hristov}, Boyan A. and {Kocevski}, Daniel and {Lysenko}, Alexandra L. and {Mailyan}, Bagrat and {Mangan}, Joseph and {McBreen}, Sheila and {Racusin}, Judith and {Ridnaia}, Anna and {Roberts}, Oliver J. and {Ulanov}, Mikhail and {Veres}, Peter and {Wilson-Hodge}, Colleen A. and {Wood}, Joshua},
        title = "{GRB 221009A: The Boat}",
      journal = {\apjl},
         year = 2023,
        month = mar,
       volume = {946},
       number = {1},
          eid = {L31},
        pages = {L31},
          doi = {10.3847/2041-8213/acc39c},
archivePrefix = {arXiv},
       eprint = {2302.14037},
 primaryClass = {astro-ph.HE},
       adsurl = {https://ui.adsabs.harvard.edu/abs/2023ApJ...946L..31B}
}

@ARTICLE{vanVelzen2019,
       author = {{van Velzen}, Sjoert and {Stone}, Nicholas C. and {Metzger}, Brian D. and {Gezari}, Suvi and {Brown}, Thomas M. and {Fruchter}, Andrew S.},
        title = "{Late-time UV Observations of Tidal Disruption Flares Reveal Unobscured, Compact Accretion Disks}",
      journal = {\apj},
         year = 2019,
        month = jun,
       volume = {878},
       number = {2},
          eid = {82},
        pages = {82},
          doi = {10.3847/1538-4357/ab1844},
archivePrefix = {arXiv},
       eprint = {1809.00003},
 primaryClass = {astro-ph.HE},
       adsurl = {https://ui.adsabs.harvard.edu/abs/2019ApJ...878...82V}
}

@MISC{Levan_HSTproposal,
       author = {{Levan}, Andrew James and {D'Elia}, Valerio and {De Pasquale}, Massimiliano and {Dimple}, Dimple and {Eyles-Ferris}, Robert and {Izzo}, Luca and {Jonker}, Peter G. and {Kaper}, Lex and {Laskar}, Tanmoy and {Le Floc'h}, Emeric and {Martin-Carrillo}, Antonio and {O'Brien}, Paul Thomas and {Piranomonte}, Silvia and {Pugliese}, Giovanna and {Schady}, Patricia and {Schneider}, Benjamin and {Tanvir}, Nial Rahil and {Vergani}, Susanna and {Xu}, Dong},
        title = "{Galactic or extragalactic? Unravelling the extraordinary GRB 250702BDE}",
 howpublished = {HST Proposal. Cycle 32, ID. \#17988},
         year = 2025,
        month = jul,
        pages = {17988},
       adsurl = {https://ui.adsabs.harvard.edu/abs/2025hst..prop17988L}
}

@ARTICLE{Levan2025_BDE,
       author = {{Levan}, Andrew J. and {Martin-Carrillo}, Antonio and {Laskar}, Tanmoy and {Eyles-Ferris}, Rob A.~J. and {Sneppen}, Albert and {Ravasio}, Maria Edvige and {Rastinejad}, Jillian C. and {Bright}, Joe S. and {Carotenuto}, Francesco and {Chrimes}, Ashley A. and {Corcoran}, Gregory and {Gompertz}, Benjamin P. and {Jonker}, Peter G. and {Lamb}, Gavin P. and {Malesani}, Daniele B. and {Saccardi}, Andrea and {S{\'a}nchez-Sierras}, Javier and {Schneider}, Benjamin and {Schulze}, Steve and {Tanvir}, Nial R. and {Vergani}, Susanna D. and {Watson}, Darach and {An}, Jie and {Bauer}, Franz E. and {Campana}, Sergio and {Cotter}, Laura and {van Dalen}, Joyce N.~D. and {D'Elia}, Valerio and {De Pasquale}, Massimiliano and {de Ugarte Postigo}, Antonio and {Dimple} and {Hartmann}, Dieter H. and {Hjorth}, Jens and {Izzo}, Luca and {Jakobsson}, P{\'a}ll and {Kumar}, Amit and {Melandri}, Andrea and {O'Brien}, Paul and {Piranomonte}, Silvia and {Pugliese}, Giovanna and {Quirola-V{\'a}squez}, Jonathan and {Starling}, Rhaana and {Tagliaferri}, Gianpiero and {Xu}, Dong and {Wortley}, Makenzie E.},
        title = "{The Day-long, Repeating GRB 250702B: A Unique Extragalactic Transient}",
      journal = {\apjl},
         year = 2025,
        month = sep,
       volume = {990},
       number = {1},
          eid = {L28},
        pages = {L28},
          doi = {10.3847/2041-8213/adf8e1},
archivePrefix = {arXiv},
       eprint = {2507.14286},
 primaryClass = {astro-ph.HE},
       adsurl = {https://ui.adsabs.harvard.edu/abs/2025ApJ...990L..28L}
}

@ARTICLE{Hinshaw2013_WMAP9,
       author = {{Hinshaw}, G. and {Larson}, D. and {Komatsu}, E. and {Spergel}, D.~N. and {Bennett}, C.~L. and {Dunkley}, J. and {Nolta}, M.~R. and {Halpern}, M. and {Hill}, R.~S. and {Odegard}, N. and {Page}, L. and {Smith}, K.~M. and {Weiland}, J.~L. and {Gold}, B. and {Jarosik}, N. and {Kogut}, A. and {Limon}, M. and {Meyer}, S.~S. and {Tucker}, G.~S. and {Wollack}, E. and {Wright}, E.~L.},
        title = "{Nine-year Wilkinson Microwave Anisotropy Probe (WMAP) Observations: Cosmological Parameter Results}",
      journal = {\apjs},
         year = 2013,
        month = oct,
       volume = {208},
       number = {2},
          eid = {19},
        pages = {19},
          doi = {10.1088/0067-0049/208/2/19},
archivePrefix = {arXiv},
       eprint = {1212.5226},
 primaryClass = {astro-ph.CO},
       adsurl = {https://ui.adsabs.harvard.edu/abs/2013ApJS..208...19H}
}

@ARTICLE{Gompertz2025,
       author = {{Gompertz}, Benjamin P. and {Levan}, Andrew J. and {Laskar}, Tanmoy and {Schneider}, Benjamin and {Chrimes}, Ashley A. and {Martin-Carrillo}, Antonio and {Sneppen}, Albert and {O'Neill}, David and {Malesani}, Daniele B. and {Jonker}, Peter G. and {Burns}, Eric and {Corcoran}, Gregory and {Cotter}, Laura and {de Ugarte Postigo}, Antonio and {De Pasquale}, Massimiliano and {Dimple} and {Eyles-Ferris}, Rob A.~J. and {Fruchter}, Andrew and {Izzo}, Luca and {Jakobsson}, P{\'a}ll and {Lamb}, Gavin P. and {Palmerio}, Jesse T. and {Pugliese}, Giovanna and {Ravasio}, Maria Edvige and {Saccardi}, Andrea and {Salvaterra}, Ruben and {Sarin}, Nikhil and {Schulze}, Steve and {Tanvir}, Nial and {Worssam}, Isabelle and {Wortley}, Makenzie E.},
        title = "{JWST Spectroscopy of GRB 250702B: An Extremely Rare and Exceptionally Energetic Burst in a Dusty, Massive Galaxy at z = 1.036}",
      journal = {\apjl},
         year = 2026,
        month = jan,
       volume = {997},
       number = {1},
          eid = {L4},
        pages = {L4},
          doi = {10.3847/2041-8213/ae2ed9},
archivePrefix = {arXiv},
       eprint = {2509.22778},
 primaryClass = {astro-ph.HE},
       adsurl = {https://ui.adsabs.harvard.edu/abs/2026ApJ...997L...4G}
}

@ARTICLE{Leja2017,
       author = {{Leja}, Joel and {Johnson}, Benjamin D. and {Conroy}, Charlie and {van Dokkum}, Pieter G. and {Byler}, Nell},
        title = "{Deriving Physical Properties from Broadband Photometry with Prospector: Description of the Model and a Demonstration of its Accuracy Using 129 Galaxies in the Local Universe}",
      journal = {\apj},
         year = 2017,
        month = mar,
       volume = {837},
       number = {2},
          eid = {170},
        pages = {170},
          doi = {10.3847/1538-4357/aa5ffe},
archivePrefix = {arXiv},
       eprint = {1609.09073},
 primaryClass = {astro-ph.GA},
       adsurl = {https://ui.adsabs.harvard.edu/abs/2017ApJ...837..170L}
}

@ARTICLE{Johnson2021,
       author = {{Johnson}, Benjamin D. and {Leja}, Joel and {Conroy}, Charlie and {Speagle}, Joshua S.},
        title = "{Stellar Population Inference with Prospector}",
      journal = {\apjs},
         year = 2021,
        month = jun,
       volume = {254},
       number = {2},
          eid = {22},
        pages = {22},
          doi = {10.3847/1538-4365/abef67},
archivePrefix = {arXiv},
       eprint = {2012.01426},
 primaryClass = {astro-ph.GA},
       adsurl = {https://ui.adsabs.harvard.edu/abs/2021ApJS..254...22J}
}

@software{Johnson2019,
       author = {{Johnson}, Benjamin D. and {Leja}, Joel L. and {Conroy}, Charlie and {Speagle}, Joshua S.},
        title = "{Prospector: Stellar population inference from spectra and SEDs}",
 howpublished = {Astrophysics Source Code Library, record ascl:1905.025},
         year = 2019,
        month = may,
          eid = {ascl:1905.025},
archivePrefix = {ascl},
       eprint = {1905.025},
       adsurl = {https://ui.adsabs.harvard.edu/abs/2019ascl.soft05025J}
}

@ARTICLE{Sears2025,
       author = {{Sears}, Huei and {Chornock}, Ryan and {Blanchard}, Peter K. and {Margutti}, Raffaella and {Villar}, V. Ashley and {Pierel}, Justin and {Vallely}, Patrick J. and {Alexander}, Kate D. and {Berger}, Edo and {Eftekhari}, Tarraneh and {Jacobson-Gal{\'a}n}, Wynn V. and {Laskar}, Tanmoy and {LeBaron}, Natalie and {Metzger}, Brian D. and {Milisavljevic}, Dan},
        title = "{Late-time HST and JWST Observations of GRB 221009A: Evidence for a Break in the Light Curve at 50 days}",
      journal = {\apj},
         year = 2025,
        month = may,
       volume = {984},
       number = {2},
          eid = {196},
        pages = {196},
          doi = {10.3847/1538-4357/adc306},
archivePrefix = {arXiv},
       eprint = {2412.02663},
 primaryClass = {astro-ph.HE},
       adsurl = {https://ui.adsabs.harvard.edu/abs/2025ApJ...984..196S}
}

@ARTICLE{OConnor2025,
       author = {{O'Connor}, Brendan and {Gill}, Ramandeep and {DeLaunay}, James and {Hare}, Jeremy and {Pasham}, Dheeraj and {Coughlin}, Eric R. and {Bandopadhyay}, Ananya and {Anumarlapudi}, Akash and {Paz Beniamini} and {Granot}, Jonathan and {Andreoni}, Igor and {Carney}, Jonathan and {Moss}, Michael J. and {G{\"o}{\u{g}}{\"u}{\textcommabelow s}}, Ersin and {Kennea}, Jamie A. and {Busmann}, Malte and {Dichiara}, Simone and {Freeburn}, James and {Gruen}, Daniel and {Hall}, Xander J. and {Palmese}, Antonella and {Parsotan}, Tyler and {Ronchini}, Samuele and {Tohuvavohu}, Aaron and {Williams}, Maia A.},
        title = "{Comprehensive X-Ray Observations of the Exceptional Ultralong X-Ray and Gamma-Ray Transient GRB 250702B with Swift, NuSTAR, and Chandra: Insights from the X-Ray Afterglow Properties}",
      journal = {\apjl},
         year = 2025,
        month = nov,
       volume = {994},
       number = {1},
          eid = {L17},
        pages = {L17},
          doi = {10.3847/2041-8213/ae1741},
archivePrefix = {arXiv},
       eprint = {2509.22787},
 primaryClass = {astro-ph.HE},
       adsurl = {https://ui.adsabs.harvard.edu/abs/2025ApJ...994L..17O}
}

@ARTICLE{Carney2025,
       author = {{Carney}, Jonathan and {Andreoni}, Igor and {O'Connor}, Brendan and {Freeburn}, James and {Skobe}, Hannah and {Westcott}, Lewi and {Busmann}, Malte and {Palmese}, Antonella and {Hall}, Xander J. and {Gill}, Ramandeep and {Beniamini}, Paz and {Coughlin}, Eric R. and {Kilpatrick}, Charles D. and {Anumarlapudi}, Akash and {Law}, Nicholas M. and {Corbett}, Hank and {Ahumada}, Tomas and {Chen}, Ping and {Conselice}, Christopher and {Damke}, Guillermo and {Das}, Kaustav K. and {Gal-Yam}, Avishay and {Gruen}, Daniel and {Heathcote}, Steve and {Hu}, Lei and {Karambelkar}, Viraj and {Kasliwal}, Mansi and {Labrie}, Kathleen and {Pasham}, Dheeraj and {Riffeser}, Arno and {Schmidt}, Michael and {Sharma}, Kritti and {Wilke}, Silona and {Zang}, Weicheng},
        title = "{Optical/Infrared Observations of the Extraordinary GRB 250702B: A Highly Obscured Afterglow in a Massive Galaxy Consistent with Multiple Possible Progenitors}",
      journal = {\apjl},
         year = 2025,
        month = dec,
       volume = {994},
       number = {2},
          eid = {L46},
        pages = {L46},
          doi = {10.3847/2041-8213/ae1d67},
archivePrefix = {arXiv},
       eprint = {2509.22784},
 primaryClass = {astro-ph.HE},
       adsurl = {https://ui.adsabs.harvard.edu/abs/2025ApJ...994L..46C}
}

@ARTICLE{Levan2016,
       author = {{Levan}, A.~J. and {Tanvir}, N.~R. and {Brown}, G.~C. and {Metzger}, B.~D. and {Page}, K.~L. and {Cenko}, S.~B. and {O'Brien}, P.~T. and {Lyman}, J.~D. and {Wiersema}, K. and {Stanway}, E.~R. and {Fruchter}, A.~S. and {Perley}, D.~A. and {Bloom}, J.~S.},
        title = "{Late Time Multi-wavelength Observations of Swift J1644+5734: A Luminous Optical/IR Bump and Quiescent X-Ray Emission}",
      journal = {\apj},
         year = 2016,
        month = mar,
       volume = {819},
       number = {1},
          eid = {51},
        pages = {51},
          doi = {10.3847/0004-637X/819/1/51},
archivePrefix = {arXiv},
       eprint = {1509.08945},
 primaryClass = {astro-ph.HE},
       adsurl = {https://ui.adsabs.harvard.edu/abs/2016ApJ...819...51L}
}

@ARTICLE{Kann2018,
       author = {{Kann}, D.~A. and {Schady}, P. and {Olivares}, E.~F. and {Klose}, S. and {Rossi}, A. and {Perley}, D.~A. and {Zhang}, B. and {Kr{\"u}hler}, T. and {Greiner}, J. and {Nicuesa Guelbenzu}, A. and {Elliott}, J. and {Knust}, F. and {Cano}, Z. and {Filgas}, R. and {Pian}, E. and {Mazzali}, P. and {Fynbo}, J.~P.~U. and {Leloudas}, G. and {Afonso}, P.~M.~J. and {Delvaux}, C. and {Graham}, J.~F. and {Rau}, A. and {Schmidl}, S. and {Schulze}, S. and {Tanga}, M. and {Updike}, A.~C. and {Varela}, K.},
        title = "{The optical/NIR afterglow of GRB 111209A: Complex yet not unprecedented}",
      journal = {\aap},
         year = 2018,
        month = oct,
       volume = {617},
          eid = {A122},
        pages = {A122},
          doi = {10.1051/0004-6361/201731292},
archivePrefix = {arXiv},
       eprint = {1706.00601},
 primaryClass = {astro-ph.HE},
       adsurl = {https://ui.adsabs.harvard.edu/abs/2018A&A...617A.122K}
}

@article{Conroy2010FSPS:Synthesis,
    title = {{FSPS: Flexible Stellar Population Synthesis}},
    year = {2010},
    journal = {ascl},
    author = {Conroy, Charlie and Gunn, James E. and Conroy, Charlie and Gunn, James E.},
    pages = {ascl:1010.043},
    url = {https://ui.adsabs.harvard.edu/abs/2010ascl.soft10043C/abstract}
}

@article{JohnsonDfm/python-fsps:V0.4.7,
    title = {{dfm/python-fsps: v0.4.7}},
    year = {2024},
    author = {Johnson, Ben and Foreman-Mackey, Dan and Sick, Jonathan and Leja, Joel and Walmsley, Mike and Tollerud, Erik and Leung, Henry and Scott, Spencer and Park, Minjung},
    url = {https://zenodo.org/records/12447779},
    doi = {10.5281/ZENODO.12447779}
}

@article{Calzetti2000TheGalaxies,
    title = {{The Dust Content and Opacity of Actively Star‐forming Galaxies}},
    year = {2000},
    journal = {The Astrophysical Journal},
    author = {Calzetti, Daniela and Armus, Lee and Bohlin, Ralph C. and Kinney, Anne L. and Koornneef, Jan and Storchi‐Bergmann, Thaisa},
    number = {2},
    month = {4},
    pages = {682--695},
    volume = {533},
    publisher = {American Astronomical Society},
    url = {https://ui.adsabs.harvard.edu/abs/2000ApJ...533..682C/abstract},
    doi = {10.1086/308692},
    issn = {0004-637X},
    arxivId = {astro-ph/9911459}
}

@article{Gallazzi2005TheUniverse,
    title = {{The ages and metallicities of galaxies in the local universe}},
    year = {2005},
    journal = {Monthly Notices of the Royal Astronomical Society},
    author = {Gallazzi, Anna and Charlot, Stéphane and Brinchmann, Jarle and White, Simon D.M. and Tremonti, Christy A.},
    number = {1},
    month = {9},
    pages = {41--58},
    volume = {362},
    publisher = {Blackwell Publishing Ltd},
    url = {https://ui.adsabs.harvard.edu/abs/2005MNRAS.362...41G/abstract},
    doi = {10.1111/j.1365-2966.2005.09321.x},
    issn = {00358711},
    arxivId = {astro-ph/0506539}
}

@article{Leja2019AnSurvey,
    title = {{An Older, More Quiescent Universe from Panchromatic SED Fitting of the 3D-HST Survey}},
    year = {2019},
    journal = {The Astrophysical Journal},
    author = {Leja, Joel and Johnson, Benjamin D. and Conroy, Charlie and Dokkum, Pieter van and Speagle, Joshua S. and Brammer, Gabriel and Momcheva, Ivelina and Skelton, Rosalind and Whitaker, Katherine E. and Franx, Marijn and Nelson, Erica J.},
    number = {2},
    month = {6},
    pages = {140},
    volume = {877},
    publisher = {IOP Publishing},
    url = {https://iopscience.iop.org/article/10.3847/1538-4357/ab1d5a https://iopscience.iop.org/article/10.3847/1538-4357/ab1d5a/meta},
    doi = {10.3847/1538-4357/AB1D5A},
    issn = {0004-637X},
    arxivId = {1812.05608}
}

@article{Steidel2016RECONCILINGGALAXIES,
    title = {{RECONCILING THE STELLAR AND NEBULAR SPECTRA OF HIGH-REDSHIFT GALAXIES*}},
    year = {2016},
    journal = {The Astrophysical Journal},
    author = {Steidel, Charles C. and Strom, Allison L. and Pettini, Max and Rudie, Gwen C. and Reddy, Naveen A. and Trainor, Ryan F.},
    number = {2},
    month = {8},
    pages = {159},
    volume = {826},
    publisher = {American Astronomical Society},
    url = {https://ui.adsabs.harvard.edu/abs/2016ApJ...826..159S/abstract},
    doi = {10.3847/0004-637x/826/2/159},
    issn = {0004-637X},
    arxivId = {1605.07186}
}

@article{Shapley:2015,
    title = {{The MOSDEF Survey: Excitation Properties of z {\{}{\textbackslash}tilde{\}} 2.3 Star-forming Galaxies}},
    year = {2015},
    journal = {ApJ},
    author = {Shapley, A.~E. and Reddy, N.~A. and Kriek, M and Freeman, W.~R. and Sanders, R.~L. and Siana, B and Coil, A.~L. and Mobasher, B and Shivaei, I and Price, S.~H. and de Groot, L},
    month = {3},
    pages = {88},
    volume = {801},
    doi = {10.1088/0004-637X/801/2/88},
    arxivId = {1409.7071}
}

@article{Leja2019HowModels,
    title = {{How to Measure Galaxy Star Formation Histories. II. Nonparametric Models}},
    year = {2019},
    journal = {The Astrophysical Journal},
    author = {Leja, Joel and Carnall, Adam C. and Johnson, Benjamin D. and Conroy, Charlie and Speagle, Joshua S.},
    number = {1},
    month = {4},
    pages = {3},
    volume = {876},
    publisher = {IOP Publishing},
    url = {https://iopscience.iop.org/article/10.3847/1538-4357/ab133c https://iopscience.iop.org/article/10.3847/1538-4357/ab133c/meta},
    doi = {10.3847/1538-4357/AB133C},
    issn = {0004-637X},
    arxivId = {1811.03637}
}

@article{Gordon2024, doi = {10.21105/joss.07023}, url = {https://doi.org/10.21105/joss.07023}, year = {2024}, publisher = {The Open Journal}, volume = {9}, number = {100}, pages = {7023}, author = {Gordon, Karl D.}, title = {dust_extinction: Interstellar Dust Extinction Models}, journal = {Journal of Open Source Software} }

@ARTICLE{Gordon2023,
       author = {{Gordon}, Karl D. and {Clayton}, Geoffrey C. and {Decleir}, Marjorie and {Fitzpatrick}, E.~L. and {Massa}, Derck and {Misselt}, Karl A. and {Tollerud}, Erik J.},
        title = "{One Relation for All Wavelengths: The Far-ultraviolet to Mid-infrared Milky Way Spectroscopic R(V)-dependent Dust Extinction Relationship}",
      journal = {\apj},
         year = 2023,
        month = jun,
       volume = {950},
       number = {2},
          eid = {86},
        pages = {86},
          doi = {10.3847/1538-4357/accb59},
archivePrefix = {arXiv},
       eprint = {2304.01991},
 primaryClass = {astro-ph.GA},
       adsurl = {https://ui.adsabs.harvard.edu/abs/2023ApJ...950...86G}
}

@ARTICLE{Gordon2009,
       author = {{Gordon}, Karl D. and {Cartledge}, Stefan and {Clayton}, Geoffrey C.},
        title = "{FUSE Measurements of Far-Ultraviolet Extinction. III. The Dependence on R(V) and Discrete Feature Limits from 75 Galactic Sightlines}",
      journal = {\apj},
         year = 2009,
        month = nov,
       volume = {705},
       number = {2},
        pages = {1320-1335},
          doi = {10.1088/0004-637X/705/2/1320},
archivePrefix = {arXiv},
       eprint = {0909.3087},
 primaryClass = {astro-ph.GA},
       adsurl = {https://ui.adsabs.harvard.edu/abs/2009ApJ...705.1320G}
}

@ARTICLE{Fitzpatrick2019,
       author = {{Fitzpatrick}, E.~L. and {Massa}, Derck and {Gordon}, Karl D. and {Bohlin}, Ralph and {Clayton}, Geoffrey C.},
        title = "{An Analysis of the Shapes of Interstellar Extinction Curves. VII. Milky Way Spectrophotometric Optical-through-ultraviolet Extinction and Its R-dependence}",
      journal = {\apj},
         year = 2019,
        month = dec,
       volume = {886},
       number = {2},
          eid = {108},
        pages = {108},
          doi = {10.3847/1538-4357/ab4c3a},
archivePrefix = {arXiv},
       eprint = {1910.08852},
 primaryClass = {astro-ph.GA},
       adsurl = {https://ui.adsabs.harvard.edu/abs/2019ApJ...886..108F}
}

@ARTICLE{Gordon2021,
       author = {{Gordon}, Karl D. and {Misselt}, Karl A. and {Bouwman}, Jeroen and {Clayton}, Geoffrey C. and {Decleir}, Marjorie and {Hines}, Dean C. and {Pendleton}, Yvonne and {Rieke}, George and {Smith}, J.~D.~T. and {Whittet}, D.~C.~B.},
        title = "{Milky Way Mid-Infrared Spitzer Spectroscopic Extinction Curves: Continuum and Silicate Features}",
      journal = {\apj},
         year = 2021,
        month = jul,
       volume = {916},
       number = {1},
          eid = {33},
        pages = {33},
          doi = {10.3847/1538-4357/ac00b7},
archivePrefix = {arXiv},
       eprint = {2105.05087},
 primaryClass = {astro-ph.GA},
       adsurl = {https://ui.adsabs.harvard.edu/abs/2021ApJ...916...33G}
}

@ARTICLE{Decleir2022,
       author = {{Decleir}, Marjorie and {Gordon}, Karl D. and {Andrews}, Jennifer E. and {Clayton}, Geoffrey C. and {Cushing}, Michael C. and {Misselt}, Karl A. and {Pendleton}, Yvonne and {Rayner}, John and {Vacca}, William D. and {Whittet}, D.~C.~B.},
        title = "{SpeX Near-infrared Spectroscopic Extinction Curves in the Milky Way}",
      journal = {\apj},
         year = 2022,
        month = may,
       volume = {930},
       number = {1},
          eid = {15},
        pages = {15},
          doi = {10.3847/1538-4357/ac5dbe},
archivePrefix = {arXiv},
       eprint = {2204.13716},
 primaryClass = {astro-ph.GA},
       adsurl = {https://ui.adsabs.harvard.edu/abs/2022ApJ...930...15D}
}

@article{Speagle2020DYNESTY:Evidences,
    title = {{DYNESTY: a dynamic nested sampling package for estimating Bayesian posteriors and evidences}},
    year = {2020},
    journal = {Monthly Notices of the Royal Astronomical Society},
    author = {Speagle, Joshua S.},
    number = {3},
    pages = {3132--3158},
    volume = {493},
    publisher = {Oxford University Press},
    url = {https://ui.adsabs.harvard.edu/abs/2020MNRAS.493.3132S/abstract},
    doi = {10.1093/MNRAS/STAA278},
    issn = {13652966},
    arxivId = {1904.02180}
}

@ARTICLE{Vreeswijk2011_grb111209Aredshift,
       author = {{Vreeswijk}, P. and {Fynbo}, J. and {Melandri}, A.},
        title = "{GRB 111209A: VLT/X-shooter redshift.}",
      journal = {GRB Coordinates Network},
         year = 2011,
        month = jan,
       volume = {12648},
        pages = {1},
       adsurl = {https://ui.adsabs.harvard.edu/abs/2011GCN.12648....1V}
}

@article{Schlafly:2011,
    title = {{Measuring Reddening with Sloan Digital Sky Survey Stellar Spectra and Recalibrating SFD}},
    year = {2011},
    journal = {ApJ},
    author = {Schlafly, E.~F. and Finkbeiner, D.~P.},
    month = {8},
    pages = {103},
    volume = {737},
    doi = {10.1088/0004-637X/737/2/103},
    arxivId = {astro-ph.GA/1012.4804}
}

@ARTICLE{SVO_cite1,
       author = {{Rodrigo}, Carlos and {Cruz}, Patricia and {Aguilar}, John F. and {Aller}, Alba and {Solano}, Enrique and {G{\'a}lvez-Ortiz}, Maria Cruz and {Jim{\'e}nez-Esteban}, Francisco and {Mas-Buitrago}, Pedro and {Bayo}, Amelia and {Cort{\'e}s-Contreras}, Miriam and {Murillo-Ojeda}, Raquel and {Bonoli}, Silvia and {Cenarro}, Javier and {Dupke}, Renato and {L{\'o}pez-Sanjuan}, Carlos and {Mar{\'\i}n-Franch}, Antonio and {de Oliveira}, Claudia Mendes and {Moles}, Mariano and {Taylor}, Keith and {Varela}, Jes{\'u}s and {Rami{\'o}}, H{\'e}ctor V{\'a}zquez},
        title = "{Photometric segregation of dwarf and giant FGK stars using the SVO Filter Profile Service and photometric tools}",
      journal = {\aap},
         year = 2024,
        month = sep,
       volume = {689},
          eid = {A93},
        pages = {A93},
          doi = {10.1051/0004-6361/202449998},
archivePrefix = {arXiv},
       eprint = {2406.03310},
 primaryClass = {astro-ph.SR},
       adsurl = {https://ui.adsabs.harvard.edu/abs/2024A&A...689A..93R}
}

@MISC{SVO_cite2,
       author = {{Rodrigo}, Carlos and {Solano}, Enrique and {Bayo}, Amelia},
        title = "{SVO Filter Profile Service Version 1.0}",
 howpublished = {IVOA Working Draft 15 October 2012},
         year = 2012,
        month = oct,
        pages = {1015},
          doi = {10.5479/ADS/bib/2012ivoa.rept.1015R},
       adsurl = {https://ui.adsabs.harvard.edu/abs/2012ivoa.rept.1015R}
}

@INPROCEEDINGS{SVO_cite3,
       author = {{Rodrigo}, C. and {Solano}, E.},
        title = "{The SVO Filter Profile Service}",
    booktitle = {XIV.0 Scientific Meeting (virtual) of the Spanish Astronomical Society},
         year = 2020,
        month = jul,
          eid = {182},
        pages = {182},
       adsurl = {https://ui.adsabs.harvard.edu/abs/2020sea..confE.182R}
}

@software{WebPlotDigitizer,
    author = {Ankit Rohatgi},
    title = {WebPlotDigitizer},
    url = {https://web.eecs.utk.edu/~dcostine/personal/PowerDeviceLib/DigiTest/index.html},
    version = {3.4},
    year = 2024
}

@software{larry_bradley_2025,
  author       = {Larry Bradley and
                  Brigitta Sip{\H o}cz and
                  Thomas Robitaille and
                  Erik Tollerud and
                  Z\`e Vin{\'{\i}}cius and
                  Christoph Deil and
                  Kyle Barbary and
                  Tom J Wilson and
                  Ivo Busko and
                  Axel Donath and
                  Hans Moritz G{\"u}nther and
                  Mihai Cara and
                  P. L. Lim and
                  Sebastian Me{\ss}linger and
                  Zach Burnett and
                  Simon Conseil and
                  Michael Droettboom and
                  Azalee Bostroem and
                  E. M. Bray and
                  Lars Andersen Bratholm and
                  William Jamieson and
                  Adam Ginsburg and
                  Geert Barentsen and
                  Matt Craig and
                  Sergio Pascual and
                  Shivangee Rathi and
                  Marshall Perrin and
                  Brett M. Morris},
  title        = {astropy/photutils: 2.2.0},
  month        = feb,
  year         = 2025,
  publisher    = {Zenodo},
  version      = {2.2.0},
  doi          = {10.5281/zenodo.14889440},
  url          = {https://doi.org/10.5281/zenodo.14889440},
  swhid        = {swh:1:dir:11159107f27a28985192ed1118b1f2055709d093
                   ;origin=https://doi.org/10.5281/zenodo.596036;visi
                   t=swh:1:snp:ae8c4a55d349d43e53cfe9ce92e678fcfe840f
                   3b;anchor=swh:1:rel:0117f67e8888adcdfc85308287dd9c
                   854b466389;path=astropy-photutils-ffb96c5
                  },
}

@ARTICLE{Blanchard2016,
       author = {{Blanchard}, Peter K. and {Berger}, Edo and {Fong}, Wen-fai},
        title = "{The Offset and Host Light Distributions of Long Gamma-Ray Bursts: A New View From HST Observations of Swift Bursts}",
      journal = {\apj},
         year = 2016,
        month = feb,
       volume = {817},
       number = {2},
          eid = {144},
        pages = {144},
          doi = {10.3847/0004-637X/817/2/144},
archivePrefix = {arXiv},
       eprint = {1509.07866},
 primaryClass = {astro-ph.HE},
       adsurl = {https://ui.adsabs.harvard.edu/abs/2016ApJ...817..144B}
}

@ARTICLE{SHOALSI,
       author = {{Perley}, D.~A. and {Kr{\"u}hler}, T. and {Schulze}, S. and {de Ugarte Postigo}, A. and {Hjorth}, J. and {Berger}, E. and {Cenko}, S.~B. and {Chary}, R. and {Cucchiara}, A. and {Ellis}, R. and {Fong}, W. and {Fynbo}, J.~P.~U. and {Gorosabel}, J. and {Greiner}, J. and {Jakobsson}, P. and {Kim}, S. and {Laskar}, T. and {Levan}, A.~J. and {Micha{\l}owski}, M.~J. and {Milvang-Jensen}, B. and {Tanvir}, N.~R. and {Th{\"o}ne}, C.~C. and {Wiersema}, K.},
        title = "{The Swift Gamma-Ray Burst Host Galaxy Legacy Survey. I. Sample Selection and Redshift Distribution}",
      journal = {\apj},
         year = 2016,
        month = jan,
       volume = {817},
       number = {1},
          eid = {7},
        pages = {7},
          doi = {10.3847/0004-637X/817/1/7},
archivePrefix = {arXiv},
       eprint = {1504.02482},
 primaryClass = {astro-ph.GA},
       adsurl = {https://ui.adsabs.harvard.edu/abs/2016ApJ...817....7P}
}

@ARTICLE{SHOALSII,
       author = {{Perley}, D.~A. and {Tanvir}, N.~R. and {Hjorth}, J. and {Laskar}, T. and {Berger}, E. and {Chary}, R. and {de Ugarte Postigo}, A. and {Fynbo}, J.~P.~U. and {Kr{\"u}hler}, T. and {Levan}, A.~J. and {Micha{\l}owski}, M.~J. and {Schulze}, S.},
        title = "{The Swift GRB Host Galaxy Legacy Survey. II. Rest-frame Near-IR Luminosity Distribution and Evidence for a Near-solar Metallicity Threshold}",
      journal = {\apj},
         year = 2016,
        month = jan,
       volume = {817},
       number = {1},
          eid = {8},
        pages = {8},
          doi = {10.3847/0004-637X/817/1/8},
archivePrefix = {arXiv},
       eprint = {1504.02479},
 primaryClass = {astro-ph.GA},
       adsurl = {https://ui.adsabs.harvard.edu/abs/2016ApJ...817....8P}
}

@ARTICLE{Eftekhari2024,
       author = {{Eftekhari}, T. and {Tchekhovskoy}, A. and {Alexander}, K.~D. and {Berger}, E. and {Chornock}, R. and {Laskar}, T. and {Margutti}, R. and {Yao}, Y. and {Cendes}, Y. and {Gomez}, S. and {Hajela}, A. and {Pasham}, D.~R.},
        title = "{Late-time X-Ray Observations of the Jetted Tidal Disruption Event AT2022cmc: The Relativistic Jet Shuts Off}",
      journal = {\apj},
         year = 2024,
        month = oct,
       volume = {974},
       number = {2},
          eid = {149},
        pages = {149},
          doi = {10.3847/1538-4357/ad72ea},
archivePrefix = {arXiv},
       eprint = {2404.10036},
 primaryClass = {astro-ph.HE},
       adsurl = {https://ui.adsabs.harvard.edu/abs/2024ApJ...974..149E}
}

@ARTICLE{Brown2015,
       author = {{Brown}, G.~C. and {Levan}, A.~J. and {Stanway}, E.~R. and {Tanvir}, N.~R. and {Cenko}, S.~B. and {Berger}, E. and {Chornock}, R. and {Cucchiaria}, A.},
        title = "{Swift J1112.2-8238: a candidate relativistic tidal disruption flare}",
      journal = {\mnras},
         year = 2015,
        month = oct,
       volume = {452},
       number = {4},
        pages = {4297-4306},
          doi = {10.1093/mnras/stv1520},
archivePrefix = {arXiv},
       eprint = {1507.03582},
 primaryClass = {astro-ph.HE},
       adsurl = {https://ui.adsabs.harvard.edu/abs/2015MNRAS.452.4297B}
}

@ARTICLE{Brown2017,
       author = {{Brown}, G.~C. and {Levan}, A.~J. and {Stanway}, E.~R. and {Kr{\"u}hler}, T. and {Tanvir}, N.~R. and {Davies}, L.~J.~M. and {Fruchter}, A. and {Cenko}, S.~B. and {Metzger}, B.~D.},
        title = "{Late-time observations of the relativistic tidal disruption flare candidate Swift J1112.2-8238}",
      journal = {\mnras},
         year = 2017,
        month = dec,
       volume = {472},
       number = {4},
        pages = {4469-4479},
          doi = {10.1093/mnras/stx2193},
archivePrefix = {arXiv},
       eprint = {1708.09668},
 primaryClass = {astro-ph.HE},
       adsurl = {https://ui.adsabs.harvard.edu/abs/2017MNRAS.472.4469B}
}

@ARTICLE{Andreoni2022,
       author = {{Andreoni}, Igor and {Coughlin}, Michael W. and {Perley}, Daniel A. and {Yao}, Yuhan and {Lu}, Wenbin and {Cenko}, S. Bradley and {Kumar}, Harsh and {Anand}, Shreya and {Ho}, Anna Y.~Q. and {Kasliwal}, Mansi M. and {de Ugarte Postigo}, Antonio and {Sagu{\'e}s-Carracedo}, Ana and {Schulze}, Steve and {Kann}, D. Alexander and {Kulkarni}, S.~R. and {Sollerman}, Jesper and {Tanvir}, Nial and {Rest}, Armin and {Izzo}, Luca and {Somalwar}, Jean J. and {Kaplan}, David L. and {Ahumada}, Tom{\'a}s and {Anupama}, G.~C. and {Auchettl}, Katie and {Barway}, Sudhanshu and {Bellm}, Eric C. and {Bhalerao}, Varun and {Bloom}, Joshua S. and {Bremer}, Michael and {Bulla}, Mattia and {Burns}, Eric and {Campana}, Sergio and {Chandra}, Poonam and {Charalampopoulos}, Panos and {Cooke}, Jeff and {D'Elia}, Valerio and {Das}, Kaustav Kashyap and {Dobie}, Dougal and {Ag{\"u}{\'\i} Fern{\'a}ndez}, Jos{\'e} Feliciano and {Freeburn}, James and {Fremling}, Cristoffer and {Gezari}, Suvi and {Goode}, Simon and {Graham}, Matthew J. and {Hammerstein}, Erica and {Karambelkar}, Viraj R. and {Kilpatrick}, Charles D. and {Kool}, Erik C. and {Krips}, Melanie and {Laher}, Russ R. and {Leloudas}, Giorgos and {Levan}, Andrew and {Lundquist}, Michael J. and {Mahabal}, Ashish A. and {Medford}, Michael S. and {Miller}, M. Coleman and {M{\"o}ller}, Anais and {Mooley}, Kunal P. and {Nayana}, A.~J. and {Nir}, Guy and {Pang}, Peter T.~H. and {Paraskeva}, Emmy and {Perley}, Richard A. and {Petitpas}, Glen and {Pursiainen}, Miika and {Ravi}, Vikram and {Ridden-Harper}, Ryan and {Riddle}, Reed and {Rigault}, Mickael and {Rodriguez}, Antonio C. and {Rusholme}, Ben and {Sharma}, Yashvi and {Smith}, I.~A. and {Stein}, Robert D. and {Th{\"o}ne}, Christina and {Tohuvavohu}, Aaron and {Valdes}, Frank and {van Roestel}, Jan and {Vergani}, Susanna D. and {Wang}, Qinan and {Zhang}, Jielai},
        title = "{A very luminous jet from the disruption of a star by a massive black hole}",
      journal = {\nat},
         year = 2022,
        month = dec,
       volume = {612},
       number = {7940},
        pages = {430-434},
          doi = {10.1038/s41586-022-05465-8},
archivePrefix = {arXiv},
       eprint = {2211.16530},
 primaryClass = {astro-ph.HE},
       adsurl = {https://ui.adsabs.harvard.edu/abs/2022Natur.612..430A}
}

@ARTICLE{Hammerstein2026,
       author = {{Hammerstein}, Erica and {Cenko}, S. Bradley and {Andreoni}, Igor and {Charalampopoulos}, Panos and {Chornock}, Ryan and {Margutti}, Raffaella and {O'Connor}, Brendan and {Schulze}, Steve and {Sollerman}, Jesper and {Barway}, Sudhanshu and {Bhalerao}, Varun and {Anupama}, G.~C. and {Kumar}, Harsh and {Marini}, Ester and {Paris}, Diego and {Perley}, Daniel A. and {Rossi}, Andrea and {Yao}, Yuhan},
        title = "{The Jetted Tidal Disruption Event AT 2022cmc: Investigating Connections to the Optical Tidal Disruption Event Population and Spectral Subclasses through Late-time Follow-up}",
      journal = {\apj},
         year = 2026,
        month = jan,
       volume = {996},
       number = {2},
          eid = {143},
        pages = {143},
          doi = {10.3847/1538-4357/ae1838},
archivePrefix = {arXiv},
       eprint = {2506.08250},
 primaryClass = {astro-ph.HE},
       adsurl = {https://ui.adsabs.harvard.edu/abs/2026ApJ...996..143H}
}

@ARTICLE{Cenko2012,
       author = {{Cenko}, S. Bradley and {Krimm}, Hans A. and {Horesh}, Assaf and {Rau}, Arne and {Frail}, Dale A. and {Kennea}, Jamie A. and {Levan}, Andrew J. and {Holland}, Stephen T. and {Butler}, Nathaniel R. and {Quimby}, Robert M. and {Bloom}, Joshua S. and {Filippenko}, Alexei V. and {Gal-Yam}, Avishay and {Greiner}, Jochen and {Kulkarni}, S.~R. and {Ofek}, Eran O. and {Olivares E.}, Felipe and {Schady}, Patricia and {Silverman}, Jeffrey M. and {Tanvir}, Nial R. and {Xu}, Dong},
        title = "{Swift J2058.4+0516: Discovery of a Possible Second Relativistic Tidal Disruption Flare?}",
      journal = {\apj},
         year = 2012,
        month = jul,
       volume = {753},
       number = {1},
          eid = {77},
        pages = {77},
          doi = {10.1088/0004-637X/753/1/77},
archivePrefix = {arXiv},
       eprint = {1107.5307},
 primaryClass = {astro-ph.HE},
       adsurl = {https://ui.adsabs.harvard.edu/abs/2012ApJ...753...77C}
}

@ARTICLE{Pasham2015,
       author = {{Pasham}, Dheeraj R. and {Cenko}, S. Bradley and {Levan}, Andrew J. and {Bower}, Geoffrey C. and {Horesh}, Assaf and {Brown}, Gregory C. and {Dolan}, Stephen and {Wiersema}, Klaas and {Filippenko}, Alexei V. and {Fruchter}, Andrew S. and {Greiner}, Jochen and {O'Brien}, Paul T. and {Page}, Kim L. and {Rau}, Arne and {Tanvir}, Nial R.},
        title = "{A Multiwavelength Study of the Relativistic Tidal Disruption Candidate Swift J2058.4+0516 at Late Times}",
      journal = {\apj},
         year = 2015,
        month = may,
       volume = {805},
       number = {1},
          eid = {68},
        pages = {68},
          doi = {10.1088/0004-637X/805/1/68},
archivePrefix = {arXiv},
       eprint = {1502.01345},
 primaryClass = {astro-ph.HE},
       adsurl = {https://ui.adsabs.harvard.edu/abs/2015ApJ...805...68P}
}

@ARTICLE{Bertin1996,
       author = {{Bertin}, E. and {Arnouts}, S.},
        title = "{SExtractor: Software for source extraction.}",
      journal = {\aaps},
         year = 1996,
        month = jun,
       volume = {117},
        pages = {393-404},
          doi = {10.1051/aas:1996164},
       adsurl = {https://ui.adsabs.harvard.edu/abs/1996A&AS..117..393B}
}

@INPROCEEDINGS{Bertin2002,
       author = {{Bertin}, Emmanuel and {Mellier}, Yannick and {Radovich}, Mario and {Missonnier}, Gilles and {Didelon}, Pierre and {Morin}, Bertrand},
        title = "{The TERAPIX Pipeline}",
    booktitle = {Astronomical Data Analysis Software and Systems XI},
         year = 2002,
       editor = {{Bohlender}, David A. and {Durand}, Daniel and {Handley}, Thomas H.},
       series = {Astronomical Society of the Pacific Conference Series},
       volume = {281},
        month = jan,
        pages = {228},
       adsurl = {https://ui.adsabs.harvard.edu/abs/2002ASPC..281..228B}
}

@software{Becker2015,
       author = {{Becker}, Andrew},
        title = "{HOTPANTS: High Order Transform of PSF ANd Template Subtraction}",
 howpublished = {Astrophysics Source Code Library, record ascl:1504.004},
         year = 2015,
        month = apr,
          eid = {ascl:1504.004},
archivePrefix = {ascl},
       eprint = {1504.004},
       adsurl = {https://ui.adsabs.harvard.edu/abs/2015ascl.soft04004B}
}

@ARTICLE{MarguttiandChornock2021,
       author = {{Margutti}, Raffaella and {Chornock}, Ryan},
        title = "{First Multimessenger Observations of a Neutron Star Merger}",
      journal = {\araa},
         year = 2021,
        month = sep,
       volume = {59},
        pages = {155-202},
          doi = {10.1146/annurev-astro-112420-030742},
archivePrefix = {arXiv},
       eprint = {2012.04810},
 primaryClass = {astro-ph.HE},
       adsurl = {https://ui.adsabs.harvard.edu/abs/2021ARA&A..59..155M}
}

@ARTICLE{WoosleyandBloom2006,
       author = {{Woosley}, S.~E. and {Bloom}, J.~S.},
        title = "{The Supernova Gamma-Ray Burst Connection}",
      journal = {\araa},
         year = 2006,
        month = sep,
       volume = {44},
       number = {1},
        pages = {507-556},
          doi = {10.1146/annurev.astro.43.072103.150558},
archivePrefix = {arXiv},
       eprint = {astro-ph/0609142},
 primaryClass = {astro-ph},
       adsurl = {https://ui.adsabs.harvard.edu/abs/2006ARA&A..44..507W}
}

@ARTICLE{Kann2019,
       author = {{Kann}, D.~A. and {Schady}, P. and {Olivares E.}, F. and {Klose}, S. and {Rossi}, A. and {Perley}, D.~A. and {Kr{\"u}hler}, T. and {Greiner}, J. and {Nicuesa Guelbenzu}, A. and {Elliott}, J. and {Knust}, F. and {Filgas}, R. and {Pian}, E. and {Mazzali}, P. and {Fynbo}, J.~P.~U. and {Leloudas}, G. and {Afonso}, P.~M.~J. and {Delvaux}, C. and {Graham}, J.~F. and {Rau}, A. and {Schmidl}, S. and {Schulze}, S. and {Tanga}, M. and {Updike}, A.~C. and {Varela}, K.},
        title = "{Highly luminous supernovae associated with gamma-ray bursts. I. GRB 111209A/SN 2011kl in the context of stripped-envelope and superluminous supernovae}",
      journal = {\aap},
         year = 2019,
        month = apr,
       volume = {624},
          eid = {A143},
        pages = {A143},
          doi = {10.1051/0004-6361/201629162},
archivePrefix = {arXiv},
       eprint = {1606.06791},
 primaryClass = {astro-ph.HE},
       adsurl = {https://ui.adsabs.harvard.edu/abs/2019A&A...624A.143K}
}

@ARTICLE{Levan2014,
       author = {{Levan}, A.~J. and {Tanvir}, N.~R. and {Starling}, R.~L.~C. and {Wiersema}, K. and {Page}, K.~L. and {Perley}, D.~A. and {Schulze}, S. and {Wynn}, G.~A. and {Chornock}, R. and {Hjorth}, J. and {Cenko}, S.~B. and {Fruchter}, A.~S. and {O'Brien}, P.~T. and {Brown}, G.~C. and {Tunnicliffe}, R.~L. and {Malesani}, D. and {Jakobsson}, P. and {Watson}, D. and {Berger}, E. and {Bersier}, D. and {Cobb}, B.~E. and {Covino}, S. and {Cucchiara}, A. and {de Ugarte Postigo}, A. and {Fox}, D.~B. and {Gal-Yam}, A. and {Goldoni}, P. and {Gorosabel}, J. and {Kaper}, L. and {Kr{\"u}hler}, T. and {Karjalainen}, R. and {Osborne}, J.~P. and {Pian}, E. and {S{\'a}nchez-Ram{\'\i}rez}, R. and {Schmidt}, B. and {Skillen}, I. and {Tagliaferri}, G. and {Th{\"o}ne}, C. and {Vaduvescu}, O. and {Wijers}, R.~A.~M.~J. and {Zauderer}, B.~A.},
        title = "{A New Population of Ultra-long Duration Gamma-Ray Bursts}",
      journal = {\apj},
         year = 2014,
        month = jan,
       volume = {781},
       number = {1},
          eid = {13},
        pages = {13},
          doi = {10.1088/0004-637X/781/1/13},
archivePrefix = {arXiv},
       eprint = {1302.2352},
 primaryClass = {astro-ph.HE},
       adsurl = {https://ui.adsabs.harvard.edu/abs/2014ApJ...781...13L}
}

@ARTICLE{Leung2026_grb220607a,
       author = {{Leung}, James K. and {Salafia}, Om Sharan and {Spingola}, Cristiana and {Ghirlanda}, Giancarlo and {Giarratana}, Stefano and {Giroletti}, Marcello and {Reynolds}, Cormac and {Wang}, Ziteng and {An}, Tao and {Deller}, Adam and {Drout}, Maria R. and {Horesh}, Assaf and {Kaplan}, David L. and {Lenc}, Emil and {Murphy}, Tara and {Perez-Torres}, Miguel and {Rhodes}, Lauren},
        title = "{The Radio Afterglow of the Ultralong GRB 220627A}",
      journal = {\apj},
         year = 2026,
        month = jan,
       volume = {996},
       number = {1},
          eid = {22},
        pages = {22},
          doi = {10.3847/1538-4357/ae1956},
archivePrefix = {arXiv},
       eprint = {2502.13435},
 primaryClass = {astro-ph.HE},
       adsurl = {https://ui.adsabs.harvard.edu/abs/2026ApJ...996...22L}
}

@ARTICLE{goad2007_grb051117a,
       author = {{Goad}, M.~R. and {Page}, K.~L. and {Godet}, O. and {Beardmore}, A. and {Osborne}, J.~P. and {O'Brien}, P.~T. and {Starling}, R. and {Holland}, S. and {Band}, D. and {Falcone}, A. and {Gehrels}, N. and {Burrows}, D.~N. and {Nousek}, J.~A. and {Roming}, P.~W.~A. and {Moretti}, A. and {Perri}, M.},
        title = "{Swift multi-wavelength observations of the bright flaring burst GRB 051117A}",
      journal = {\aap},
         year = 2007,
        month = jun,
       volume = {468},
       number = {1},
        pages = {103-112},
          doi = {10.1051/0004-6361:20066874},
archivePrefix = {arXiv},
       eprint = {astro-ph/0612661},
 primaryClass = {astro-ph},
       adsurl = {https://ui.adsabs.harvard.edu/abs/2007A&A...468..103G}
}

@ARTICLE{Fu2024_grb211024b,
       author = {{Fu}, Shao-Yu and {Xu}, Dong and {Lei}, Wei-Hua and {de Ugarte Postigo}, Antonio and {Malesani}, Daniele B. and {Kann}, David Alexander and {Jakobsson}, P{\'a}ll and {Fynbo}, Johan P.~U. and {Maiorano}, Elisabetta and {Rossi}, Andrea and {Paris}, Diego and {Liu}, Xing and {Jiang}, Shuai-Qing and {Lu}, Tian-Hua and {An}, Jie and {Zhu}, Zi-Pei and {Gao}, Xing and {Wei}, Jian-Yan},
        title = "{GRB 211024B: An Ultra-long GRB Powered by Magnetar}",
      journal = {\apj},
         year = 2024,
        month = dec,
       volume = {977},
       number = {2},
          eid = {197},
        pages = {197},
          doi = {10.3847/1538-4357/ad8886},
archivePrefix = {arXiv},
       eprint = {2410.15162},
 primaryClass = {astro-ph.HE},
       adsurl = {https://ui.adsabs.harvard.edu/abs/2024ApJ...977..197F}
}

@ARTICLE{Virgili2013_grb091024A,
       author = {{Virgili}, F.~J. and {Mundell}, C.~G. and {Pal'shin}, V. and {Guidorzi}, C. and {Margutti}, R. and {Melandri}, A. and {Harrison}, R. and {Kobayashi}, S. and {Chornock}, R. and {Henden}, A. and {Updike}, A.~C. and {Cenko}, S.~B. and {Tanvir}, N.~R. and {Steele}, I.~A. and {Cucchiara}, A. and {Gomboc}, A. and {Levan}, A. and {Cano}, Z. and {Mottram}, C.~J. and {Clay}, N.~R. and {Bersier}, D. and {Kopa{\v{c}}}, D. and {Japelj}, J. and {Filippenko}, A.~V. and {Li}, W. and {Svinkin}, D. and {Golenetskii}, S. and {Hartmann}, D.~H. and {Milne}, P.~A. and {Williams}, G. and {O'Brien}, P.~T. and {Fox}, D.~B. and {Berger}, E.},
        title = "{GRB 091024A and the Nature of Ultra-long Gamma-Ray Bursts}",
      journal = {\apj},
         year = 2013,
        month = nov,
       volume = {778},
       number = {1},
          eid = {54},
        pages = {54},
          doi = {10.1088/0004-637X/778/1/54},
archivePrefix = {arXiv},
       eprint = {1310.0313},
 primaryClass = {astro-ph.HE},
       adsurl = {https://ui.adsabs.harvard.edu/abs/2013ApJ...778...54V}
}

@ARTICLE{Levan2015_ULGRBreview,
       author = {{Levan}, A.~J.},
        title = "{Swift discoveries of new populations of extremely long duration high energy transient}",
      journal = {Journal of High Energy Astrophysics},
         year = 2015,
        month = sep,
       volume = {7},
        pages = {44-55},
          doi = {10.1016/j.jheap.2015.05.004},
archivePrefix = {arXiv},
       eprint = {1506.03960},
 primaryClass = {astro-ph.HE},
       adsurl = {https://ui.adsabs.harvard.edu/abs/2015JHEAp...7...44L}
}

@ARTICLE{Wu2013_GRB121027A,
       author = {{Wu}, Xue-Feng and {Hou}, Shu-Jin and {Lei}, Wei-Hua},
        title = "{Giant X-Ray Bump in GRB 121027A: Evidence for Fall-back Disk Accretion}",
      journal = {\apjl},
         year = 2013,
        month = apr,
       volume = {767},
       number = {2},
          eid = {L36},
        pages = {L36},
          doi = {10.1088/2041-8205/767/2/L36},
archivePrefix = {arXiv},
       eprint = {1302.4878},
 primaryClass = {astro-ph.HE},
       adsurl = {https://ui.adsabs.harvard.edu/abs/2013ApJ...767L..36W}
}

@ARTICLE{Neights2026,
       author = {{Neights}, Eliza and {Burns}, Eric and {Fryer}, Chris L. and {Svinkin}, Dmitry and {Bala}, Suman and {Hamburg}, Rachel and {Gill}, Ramandeep and {Negro}, Michela and {Masterson}, Megan and {DeLaunay}, James and {Lawrence}, David J. and {Abrahams}, Sophie E.~D. and {Kawakubo}, Yuta and {Beniamini}, Paz and {Diget}, Christian Aa and {Frederiks}, Dmitry and {Goldsten}, John and {Goldstein}, Adam and {Hall-Smith}, Alexander D. and {Kara}, Erin and {Laird}, Alison M. and {Lamb}, Gavin P. and {Roberts}, Oliver J. and {Seeb}, Ryan and {Villar}, V. Ashley and {Airasca}, Aldana Holzmann and {Barber}, Joseph R. and {Bhat}, P. Narayana and {Bissaldi}, Elisabetta and {Briggs}, Michael S. and {Cleveland}, William H. and {Dalessi}, Sarah and {Depalo}, Davide and {Giles}, Misty M. and {Granot}, Jonathan and {Hristov}, Boyan A. and {Hui}, C. Michelle and {von Kienlin}, Andreas and {Kierans}, Carolyn and {Kocevski}, Daniel and {Lesage}, Stephen and {Lysenko}, Alexandra L. and {Mailyan}, Bagrat and {Malacaria}, Christian and {Mukherjee}, Oindabi and {Parsotan}, Tyler and {Ridnaia}, Anna and {Ronchini}, Samuele and {Scotton}, Lorenzo and {Trigg}, Aaron C. and {Tsvetkova}, Anastasia and {Ulanov}, Mikhail and {Veres}, P{\'e}ter and {Williams}, Maia and {Wilson-Hodge}, Colleen A. and {Wood}, Joshua},
        title = "{GRB 250702B: discovery of a gamma-ray burst from a black hole falling into a star}",
      journal = {\mnras},
         year = 2026,
        month = jan,
       volume = {545},
       number = {2},
          eid = {staf2019},
        pages = {staf2019},
          doi = {10.1093/mnras/staf2019},
archivePrefix = {arXiv},
       eprint = {2509.22792},
 primaryClass = {astro-ph.HE},
       adsurl = {https://ui.adsabs.harvard.edu/abs/2026MNRAS.545f2019N}
}

@ARTICLE{Li2026_EPpaperontheGRB,
       author = {{Li}, Dongyue and {Zhang}, Wenda and {Yang}, Jun and {Chen}, Jin-Hong and {Yuan}, Weimin and {Cheng}, Huaqing and {Xu}, Fan and {Shu}, Xinwen and {Shen}, Rong-Feng and {Jiang}, Ning and {Zhu}, Jiazheng and {Zhou}, Chang and {Lei}, Weihua and {Sun}, Hui and {Jin}, Chichuan and {Dai}, Lixin and {Zhang}, Bing and {Yang}, Yu-Han and {Zhang}, Wenjie and {Feng}, Hua and {Liu}, Bifang and {Zhou}, Hongyan and {Pan}, Haiwu and {Liu}, Mingjun and {Corbel}, St{\'e}phane and {Jagan}, Sitha K. and {Baglio}, Maria Cristina and {Burns}, Christopher R. and {Cangemi}, Floriane and {Chen}, Chun and {Cheng}, Yehao and {Coleiro}, Alexis and {Coti Zelati}, Francesco and {Das}, Sourya R. and {Dong}, Zhongnan and {Galbany}, Luis and {Grollimund}, Noa and {Kelson}, Daniel and {Lai}, Dong and {Li}, Xia and {Liu}, Yuan and {Marino}, Alessio and {Mockler}, Brenna and {O'Brien}, Paul and {Qiao}, Erlin and {Rea}, Nanda and {Resmi}, L. and {Rodriguez}, J{\'e}rome and {Saxton}, Richard and {Sun}, Luming and {Tao}, Lian and {Wang}, Tinggui and {Wang}, Yilong and {Wu}, Xuefeng and {Xu}, Dong and {Zhang}, Yijia and {Zhao}, Guoying and {Bao}, Congying and {Cai}, Zhiming and {Chen}, Yehai and {Chen}, Yong and {Cordier}, Bertrand and {Cui}, Chenzhou and {Cui}, Weiwei and {Fan}, Zhou and {Gao}, He and {Ghirlanda}, Giancarlo and {Guan}, Ju and {Han}, Dawei and {Hao}, Jinxin and {Hu}, Jingwei and {Huang}, Maohai and {Huang}, Yong-Feng and {Jia}, Shumei and {Jin}, Ge and {Komossa}, Stefanie and {Li}, Chengkui and {Ling}, Zhixing and {Liu}, Congzhan and {Liu}, Heyang and {Liu}, Huaqiu and {Lu}, Fangjun and {Nandra}, Kirpal and {Ness}, Jan-Uwe and {Rau}, Arne and {Sanders}, Jeremy and {Song}, Liming and {Soria}, Roberto and {Sun}, Shengli and {Sun}, Xiaojin and {Tan}, Yuyin and {Troja}, Eleonora and {Wen}, Sixiang and {Xu}, Haitao and {Xue}, Changbin and {Xue}, Yongquan and {Yin}, Yi-Han Iris and {Zhang}, Chen and {Zhang}, Shuang-Nan and {Zhang}, Yonghe},
        title = "{A fast powerful X-ray transient from possible tidal disruption of a white dwarf}",
      journal = {Science Bulletin},
         year = 2026,
        month = feb,
       volume = {71},
       number = {3},
        pages = {538-546},
          doi = {10.1016/j.scib.2025.12.050},
archivePrefix = {arXiv},
       eprint = {2509.25877},
 primaryClass = {astro-ph.HE},
       adsurl = {https://ui.adsabs.harvard.edu/abs/2026SciBu..71..538L}
}

@ARTICLE{Greiner2015,
       author = {{Greiner}, Jochen and {Mazzali}, Paolo A. and {Kann}, D. Alexander and {Kr{\"u}hler}, Thomas and {Pian}, Elena and {Prentice}, Simon and {Olivares E.}, Felipe and {Rossi}, Andrea and {Klose}, Sylvio and {Taubenberger}, Stefan and {Knust}, Fabian and {Afonso}, Paulo M.~J. and {Ashall}, Chris and {Bolmer}, Jan and {Delvaux}, Corentin and {Diehl}, Roland and {Elliott}, Jonathan and {Filgas}, Robert and {Fynbo}, Johan P.~U. and {Graham}, John F. and {Guelbenzu}, Ana Nicuesa and {Kobayashi}, Shiho and {Leloudas}, Giorgos and {Savaglio}, Sandra and {Schady}, Patricia and {Schmidl}, Sebastian and {Schweyer}, Tassilo and {Sudilovsky}, Vladimir and {Tanga}, Mohit and {Updike}, Adria C. and {van Eerten}, Hendrik and {Varela}, Karla},
        title = "{A very luminous magnetar-powered supernova associated with an ultra-long {\ensuremath{\gamma}}-ray burst}",
      journal = {\nat},
         year = 2015,
        month = jul,
       volume = {523},
       number = {7559},
        pages = {189-192},
          doi = {10.1038/nature14579},
archivePrefix = {arXiv},
       eprint = {1509.03279},
 primaryClass = {astro-ph.HE},
       adsurl = {https://ui.adsabs.harvard.edu/abs/2015Natur.523..189G}
}
\bibliographystyle{aasjournalv7}

\appendix
\section{Addendum for \texttt{Prospector} Modeling}
\label{sec:appendixCorner}

\begin{figure}
\centering
\includegraphics[width = \textwidth]{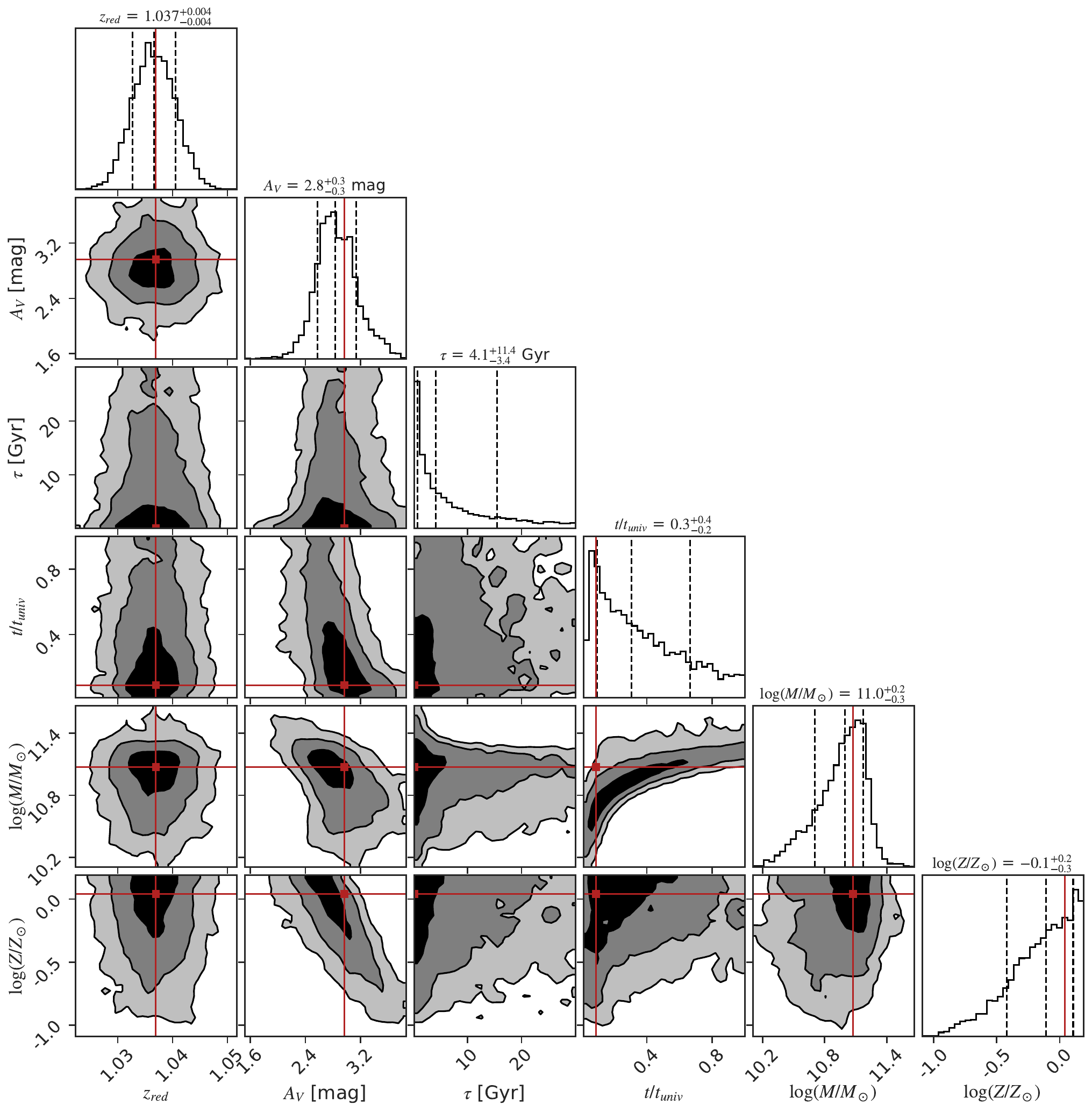}
\caption{The full corner plot showing the posterior distributions from the model. Variables are as defined in Section \ref{subsec:Proscpector}.} \label{fig:cornerplot}
\end{figure}

In Figure \ref{fig:cornerplot}, we show the posterior distributions from our \texttt{Prospector} model.

\end{document}